\documentclass[a4paper]{revtex4}

\usepackage{graphicx}
\usepackage{subeqn}
\usepackage[usenames]{color}
\usepackage{ulem}

\newcommand{\lint}[3]{\int\limits_{#1}^{#2}\!\!\mathrm{d}#3\,\,}

\begin{document}

~
\setcounter{page}{0}
\thispagestyle{empty}
\vspace{20cm}

Publishing information:

T. K\"uhn and G. S. Paraoanu, {\it Electronic and thermal sequential transport in metallic and superconducting two-junction arrays}, in {\sl Trends in nanophysics: theory, experiment, technology}, edited by V. Barsan and A. Aldea,  Engineering Materials Series, Springer-Verlag, Berlin (ISBN: 978-3-642-12069-5), pp. 99-131 (2010).

DOI: 10.1007/978-3-642-12070-1\_5

\newpage

\title{Electronic and thermal sequential transport in metallic and superconducting two-junction arrays}

\author{T.~K\"uhn}

\affiliation{NanoScience Center and Department of Physics, University of Jyv\"askyl\"a,
P.O.~Box 35 (YFL), FIN-40014 University of Jyv\"askyl\"a, Finland}

\author{G.~S.~Paraoanu}\email{paraoanu@cc.hut.fi}
\affiliation{Low Temperature Laboratory, School of Science and Technology, Aalto University, P. O. Box 15100, FI-00076 AALTO, Finland.}

\begin{abstract}

The description of transport phenomena in devices consisting of arrays of tunnel junctions, and the  experimental
confirmation of these predictions is one of the great successes of mesoscopic physics. The aim of this paper is to
give a self-consistent review of sequential transport processes in
such devices, based on the so-called "orthodox" model. We calculate numerically the current-voltage ($I$--$V$) curves, the conductance versus bias voltage ($G$--$V$) curves,
and the associated thermal transport in symmetric and asymmetric
two-junction arrays such as Coulomb-blockade thermometers (CBTs),
superconducting-insulator-normal-insulator-superconducting (SINIS) structures,
and superconducting single-electron transistors (SETs).
We investigate the behavior of these systems at the singularity-matching bias
points, the dependence of microrefrigeration
effects on the charging energy of the island, and the effect of a finite
superconducting gap on Coulomb-blockade thermometry.

\end{abstract}

\maketitle

\section{Introduction}

Quasiparticle transport processes across metallic junctions play a fundamental role in the functioning of many devices
used nowadays in mesoscopic physics. One such device is the single electron transistor (SET), invented and fabricated almost two decades ago
\cite{averinlikharev}, which has found remarkable applications as ultrasensitive charge detector \cite{electrometer} and as amplifier operating at the quantum limit \cite{quantumamplifier}. In the emerging field of quantum computing the superconducting SET has been proposed and used as a
quantum bit \cite{qubit}. With advancements in lithography techniques, this device can be fully suspended \cite{suset}, thus providing a new avenue for nanoelectromechanics.  Similar devices are currently used (in practice arrays with several junctions turn out to provide a larger signal-to-noise ratio) as Coulomb blockade (CBT) primary thermometers \cite{cbt}. Also, superconducting double-junction
systems with appropriate bias can be operated as
microcoolers \cite{microcoolers}. The functioning of these three classes of devices is based on the interplay between
two out of the three relevant energy scales: the superconducting gap, the charging energy, and the temperature.

For example, in the case of
microcoolers, the temperature and the gap are finite, and the charging energy is typically zero.
A natural question to raise is then what happens if the charging energy is no longer negligible, for
example if one wishes to miniaturize further these devices. In contradistinction, for CBTs the charging energy and the
temperature are important, and the superconducting gap is a nuisance. A solution is to suppress
the gap by using external magnetic fields, an idea which makes these temperature sensors more bulky
and risky to use near magnetic-field sensitive components. Therefore, understanding the corrections
introduced by the superconducting gap could provide an interesting alternative route, although, with present technology,
the level of control required of the gap value could be very difficult to achieve. Finally, electrometers and superconducting
SETs are operated at low temperatures, with the charging energy and the gap being dominant. However,
large charging energies are not always easy to obtain for some materials due to technological
limitations, while achieving effective very low electronic temperatures is limited by various
nonequilibrium processes.

In this article we present a unified treatment of these three devices by solving the transport
problem in the most general case, when all three energy scales are present. Our goal  is to give an eye guidance for the experimentalist working in the field, showing what are the
main characteristics  visible in the $I$--$V$s and $G$--$V$s, resulting from sequential tunneling. Both the electrical and
the thermal transport are calculated in the framework of a generalized so-called "orthodox" theory, which includes
the superconducting gap. For completeness, we offer a self-consistent review of this theory, which
has become nowadays the standard model for
describing sequential quasiparticle transport processes in these devices. We ignore Josephson effects which, for the purpose of this analysis just
add certain well-known features at certain values of the bias (the Josephson
supercurrent at zero bias and the Josephson-quasiparticle (JQP) peak at $2\Delta$ for symmetric SETs) to the
characteristic.
Higher-order tunneling phenomena
such as Andreev reflections and cotunneling are neglected as well (the former is visible only for junctions with
high transparencies, which is not our case, the latter is a small, second-order transport effect across the island).
Also the effect of the environment is not discussed here, since it depends on the specifics of the sample
(for example the coupling capacitances of the electrodes). But even stripped down
to the essentials of sequential tunneling, the physics of these devices is complex enough to justify a careful analysis, as we
attempt below. We present in detail the theory and the numerical methods that can be used to characterize
these devices. Also, a lot of interest exists nowadays in the study of hybrid structures -- in which the electrodes and the island
are made from different materials, thus having different superconducting gaps.
Various
technologies have been developed to fabricate single-electron transistors with superconducting materials
other than Al, most notably with Nb which has a much larger gap \cite{us}. We extend our analysis also to such hybrid SETs.

This paper is organized in the following way: besides the introductory part, there are two main sections,
Section \ref{sec_SIS} and Section \ref{trans}, followed by closing remarks in Section \ref{conclu}.
Section \ref{sec_SIS} discusses tunneling of electrons from one superconductor into another through
an insulating barrier. Here  we show how to calculate the currents that flow through the  junction by
using the Fermi golden rule (Subsection \ref{gen}), and  we give a comprehensive presentation of the
numerical methods used to calculate the integrals over the density of states and Fermi functions which
appear naturally in this theory (Subsection \ref{num}). In Section \ref{trans} we use these results to
describe transport phenomena in two-junction devices. We assume a quasi-equilibrium situation (electron distribution described by a Fermi function with a certain
effective temperature) for both the island and the leads. We start with a presentation of the master
equation in Subsection \ref{mmaster}. A first application of this formalism is described in Subsection \ref{cbl},
where we analyze the effect of a finite superconducting gap for the functioning of Coulomb-blockade thermometers.
Then we discuss the effect of a nonzero charging energy for superconductor-insulator-normal
metal-insulator-superconductor (SINIS) structures (Subsection \ref{sinis}) -- which normally have a
negligible charging energy. After that, in Subsection \ref{sset}, we
consider the case of superconducting single-electron transistors with either identical materials in
the leads and island (symmetric SET) or with an island with a gap different from that of the electrodes
(asymmetric SET). The effect of the gate in such transistors is further investigated in
Subsection \ref{gate}. We end this section with a calculation of expected cooling effects in
such structures (Subsection  \ref{cooling}). We delegate some of  the calculations of integrals in
Appendix \ref{sec_useful}, and of expansions in Appendix \ref{sec_expansions}.

\section{Transport in single tunnel junctions}\label{sec_SIS}

Standard lithographic techniques enable the fabrication of junctions with
almost any geometrical characteristics down to the linewidth of the
respective system. When using scanning-electron microscopes, the typical linewidths obtained are
of the order of 20 nm
and even below. More advanced methods, such as focused ion beam lithography,
have started to be used recently. The metals are evaporated in an
ultra-high vacuum evaporator, using the shadow evaporation technique. This
technique consists of  evaporation through a mask realized in the resist, at
different angles, so that some lines would overlap in a certain region, thus defining
the area of the junction. In between two
evaporations, the sample is oxidized (typically aluminum oxide), a process which results
in the formation of the insulating barrier.
\begin{figure}[t]
  \begin{center}
  \includegraphics[width=4.5cm]{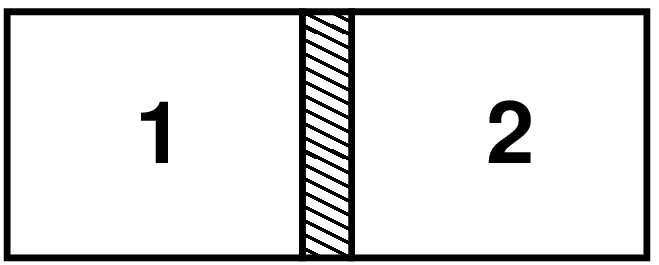} \includegraphics[width=11.5cm]{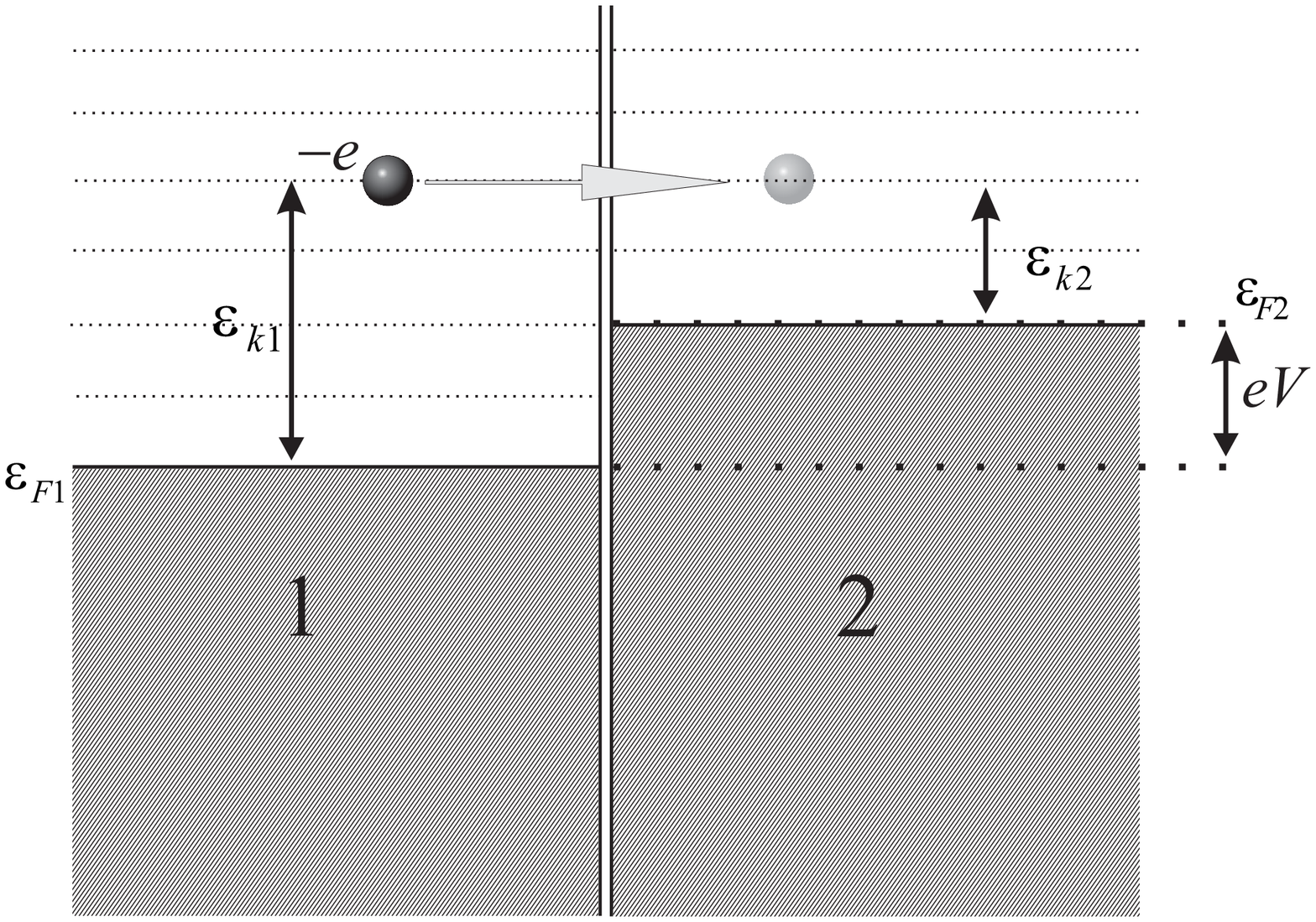}
  \end{center}
  \caption{Schematic of a tunnel junction. In the left figure, the leads are the clear areas marked with $1$ and $2$,
           while the insulating barrier is depicted by the shaded area. A schematic of the energy levels is shown for
the case of metallic electrodes.}\label{fig_junction}
\end{figure}

In this section, we assume that the area of the junction thus formed is large enough  so that capacitive effects can
be neglected (Coulomb-blockade effects are discussed in the next section). This allows us to concentrate on tunneling
effects only. The only energy scales relevant here will then be the
temperature and the tunneling matrix element.

\subsection{General theory of tunneling}
\label{gen}

Consider the case of two conductors (leads) that are connected through a thin insulating barrier. The setup
is illustrated in Fig.\ \ref{fig_junction}.
This creates effectively a potential barrier between the "left" and "right" electrodes: however electrons
can still tunnel from one side to another {\it via} the tunneling effect.
The total many-body Hamiltonian of the system comprises then two parts: one is the Hamiltonian corresponding
to the left and right electrodes (which can be either the free-electron Hamiltonian in the case of metals,
or the BCS Hamiltonian in the case of superconductors), the other is the tunneling Hamiltonian.
This Hamiltonian was first introduced by Bardeen \cite{bardeen}:
\begin{equation}
H_{T} = \sum_{k_{1}, k_{2}, \sigma} T_{k_{1},k_{2}}^{\phantom\dagger}
       c^{\dagger}_{k_{2}\sigma}c_{k_{1}\sigma}^{\phantom\dagger} + {\rm h.c.},
\end{equation}
where  $k_{1}$ and $k_{2}$ are the momentum indices corresponding to the left and right electrodes,
and $\sigma$ is the spin index (conserved under tunneling); $c^{\dagger}_{k_{1,2}\sigma}$ and $c^{\phantom\dagger}_{k_{1,2}\sigma}$
 are creation and annihilation
operators for fermions.
It is then straightforward to calculate the probability per  unit time
(the tunneling rate) that an electron residing in the left electrode would tunnel from a state $\{k_{1},\sigma \}$
(of energy $\epsilon_{k_{1}}=\hbar^{2}k_{1}^{2}/2m -\epsilon_{\rm F1}$, where $\epsilon_{\rm F1}$ is the Fermi
level of the left electrode) to a state $\{k_{2},\sigma\}$ (of energy
$\epsilon_{k_2} = \hbar^{2}k_{2}^{2}/2m -\epsilon_{\rm F2}$, where $\epsilon_{\rm F2}$ is the
Fermi energy of the right electrode), by a direct application of
the Fermi Golden rule,
\begin{equation}
P_{1\rightarrow 2} (k_{1}, k_{2})= \frac{2\pi}{\hbar} |T_{k_{1},k_{2}}|^{2}
  \rho_{k_{2}}f_{k_{1}}(1-f_{k_{2}}), \label{gr}
\end{equation}
where $f_{k_{1,2}}$ are the Fermi functions corresponding to the states $k_{1,2}$,
$\rho_{k_{2}}$ is the density of states (without the spin degrees of freedom) at $k_{2}$,
and $\delta E$ is the energy difference between the states $\{k_{2},\sigma\}$ and
$\{k_{1},\sigma \}$, $\delta E_{1\rightarrow 2}=\epsilon_{k_2}-\epsilon_{k_1}$. The energy difference $\delta E_{1\rightarrow 2}$ does not depend on  the state $k_{1}$ or $k_{2}$ since
it is due
to electrical potentials such as an external battery or to the charging of the island (as we will see below in the case of single-electron transistor) which produce a difference in the Fermi energies of the two metals, $\delta E_{1\rightarrow 2} = \epsilon_{\rm F_1}-\epsilon_{\rm F_2}$, meaning that
$\epsilon_{k_1}+\epsilon_{\rm F1} = \epsilon_{k_2}+\epsilon_{\rm F2}$, and thus ensuring  that the total energy is conserved under tunneling.

Replacing now the sum over states with a continuous integral, we can find the total tunneling rate from left to right by integrating $P_{1\rightarrow 2}$,
weighted with the density of states $\rho_{1}$,
\begin{equation}
\Gamma_{1\rightarrow 2} (\delta E_{1\rightarrow 2})= 2 \int d\epsilon_{1}\rho_{1}(\epsilon_{1})P_{1\rightarrow 2} (\epsilon_{1}, \epsilon_{2}=\epsilon_{1}+\delta E_{1\rightarrow 2} ) =
2 \int d\epsilon_{1} \int d\epsilon_{2} \rho_{1}(\epsilon_{1})P_{1\rightarrow 2} (\epsilon_{1}, \epsilon_{2}) \delta (\epsilon_{2}-\epsilon_{1}-\delta E_{1\rightarrow 2}). \label{gam}
\end{equation}
The last form of the equation, using the Dirac delta function (in no danger of confusion with the $\delta$
of $\delta E_{1\rightarrow 2}$), reflects indeed the conservation of energy: electrons of energy $\epsilon_{1}+\epsilon_{\rm F1}$
can get only to states $\epsilon_{2}+\epsilon_{\rm F2}$, if they acquire
an energy gain of $\delta E_{1\rightarrow 2}$ with respect to their Fermi levels. Inelastic processes are allowed in a more
general formulation,
which includes the degrees of freedom of the environment, in which case $\delta$ is replaced by the so-called
$P$-function \cite{ingold}.
Also, the factor of 2 appearing in front
of the integral comes from summation over the spin index.
The rate of tunneling from the right electrode to the left one,
$\Gamma_{2\rightarrow 1} (\delta E_{2\rightarrow 1})$, is calculated in a similar way.
The difference between these quantities represents the net
current of charges flowing through the junction, therefore the net electrical current is
\begin{equation}
I_{1\rightarrow 2} = -e\left[ \Gamma_{1\rightarrow 2} (\delta E_{1\rightarrow 2}) - \Gamma_{2\rightarrow 1} (\delta E_{2\rightarrow 1})\right] = -e\left[ \Gamma_{1\rightarrow 2} (\delta E_{1\rightarrow 2}) - \Gamma_{2\rightarrow 1} (-\delta E_{1\rightarrow 2})\right] \label{trr}
\end{equation}
We then get, using the expressions Eq.\ (\ref{gam}) and Eq.\ (\ref{gr}),
\begin{equation}
I_{1\rightarrow 2} = -\frac{4\pi e}{\hbar}\int d\epsilon_{1}d\epsilon_{2} |T(\epsilon_{1}, \epsilon_{2})|^{2} \rho (\epsilon_{1})
\rho (\epsilon_{2})[f(\epsilon_{1})-f(\epsilon_{2})]\delta (\epsilon_{2}-\epsilon_{1}-\delta E_{1\rightarrow 2}), \label{current}
\end{equation}
where we conveniently change all the momentum indices to continuous-energies functions.
To make progress with Eq.\ (\ref{current}), a few assumptions are needed. First of all,
already embedded in the theory above is the idea that tunneling does not significantly
change the Fermi distribution function. Nonequilibrium processes however can play an
important role, requiring a more involved treatment using Green's functions \cite{review}.
Next, the Fermi energy in most metals is of the order of a few $eV$s; this is large
when compared to the typical bias voltages at which such junctions are studied.
Then, in the case of metals, the density of states can be considered
constant around the corresponding Fermi energies $\rho (\epsilon_{1})\approx \rho (0)$,
and $\rho (\epsilon_{2})\approx \rho (0)$. For superconducting electrodes this assumption
does not hold: the density of states is divergent around the gap, and, as we will see
below, this has significant effects. It is however possible (see below) to isolate
the divergent part of the density of states as an adimensional quantity which, when multiplied
with the normal-metal density of states around the Fermi level would give an approximate but
still correct superconducting density of states. Finally, the tunneling matrix element $T$
can also be taken as energy-independent, $T (\epsilon_{1}, \epsilon_{2})=T$.
With these assumptions, the integral over energies in Eq.\ (\ref{gam}) can be performed (see \cite{ingold,schon} and Appendix \ref{sec_useful}),
using the generally useful result
\begin{equation}
\int d\epsilon f(\epsilon )[1-f(\epsilon - E)] = \frac{E}{\exp (E/k_{B}T)-1},\label{rrrt}
\end{equation}
to give for the tunneling rates
\begin{equation}
\Gamma_{1\rightarrow 2} (\delta E_{1\rightarrow 2}) = -\frac{4\pi}{\hbar} |T|^{2}\rho (\epsilon_{\rm F1})
\rho (\epsilon_{\rm F2})\frac{\delta E_{1\rightarrow 2}}{\exp (-\delta E_{1\rightarrow 2} /k_{B}T)-1} .\label{smr}
\end{equation}
For example, at zero temperature we find $\Gamma_{1\rightarrow 2} (\delta E_{1\rightarrow 2}) = (4\pi /\hbar ) |T|^{2}\rho (\epsilon_{\rm F1})
\rho (\epsilon_{\rm F2})\delta E_{1\rightarrow 2}$ for $\delta E_{1\rightarrow 2} >0$ and $\Gamma_{1\rightarrow 2} = 0$ for $\delta E_{1\rightarrow 2} \leq 0$.
The value of the current does not depend however on temperature. From Eq. (\ref{smr}) and Eq. (\ref{trr}) we find a linear dependence in $\delta E$ for the current
\begin{equation}
I_{1\rightarrow 2}  = - \frac{4\pi e}{\hbar} \rho (\epsilon_{\rm F1})\rho (\epsilon_{\rm F2}) |T|^{2} \delta{E_{1\rightarrow 2}} \label{aivi}
\end{equation}
If $\delta E_{1\rightarrow 2} = - eV$, as it is in the case of a voltage-biased single junction (see Fig. \ref{fig_junction}), we immediately
recognize Ohm's law from Eq.\ (\ref{aivi}), $I_{1\rightarrow 2} (V) = R_{12}^{-1}V$;
with the quanta of resistance defined as $R_{K} = h/e^{2} = 2\pi \hbar /e^{2} = 25.8$ ${\rm k}\Omega$ we get
\begin{equation}
\frac{1}{R_{12}} = \frac{8\pi^2}{R_{K}} \rho (\epsilon_{\rm F1})\rho (\epsilon_{\rm F2})
|T|^{2}.\label{xrt}
\end{equation}
The theory presented above looks very reasonable until one realizes -- {\it horribile dictu}! -- that
in fact the result Eq.\ (\ref{xrt}) predicts a rather absurd scaling of electrical resistance with the
dimensions of the sample. Let us look at the densities of state. The three-dimensional density of states
per volume (with spin states excluded) can be calculated using the standard text-book expression
\begin{equation}
{\cal G}^{(3D)} (E) = \frac{1}{4\pi^2} \left(\frac{2m}{\hbar^2} \right) \sqrt{E}.
\end{equation}
The Fermi energy for metals is of the order of a few eV: for example, $\epsilon_{\rm F}^{\rm (Au)} = 5.53$ eV,
$\epsilon_{\rm F}^{\rm (Ag)} = 5.94$ eV, $\epsilon_{\rm F}^{\rm (Al)} = 11.63$ eV, $\epsilon_{\rm F}^{\rm (Cu)} = 7$ eV.
At $\epsilon_{\rm F}=10$
eV we obtain for example ${\cal G}^{(3D)} (10 eV) = 0.75 \times 10^{57} J^{-1}{\rm m}^{-3}$; the Fermi wavelength at
$\epsilon_{\rm F}=10$ eV is $\lambda_{\rm F}=h/\sqrt{2m \epsilon_{\rm F}}$= 0.37 nm.

The problem now is that in order to find $\rho (\epsilon_{\rm F})$ we have to multiply the density of states per volume
${\cal G}^{3D}$ with the volume of the electrodes. This is of course absurd, as the inverse of resistance should
be linear in the junction's surface and not depend on the volume of the electrodes.  The solution to this
conundrum stems from the idea that the summation in Eq.\ (\ref{gam}) has to be done under the restriction
that transversal momentum is conserved (specular transmission): thus the correct densities of states that
appear further have to be densities of states at fixed transversal momentum \cite{harrison}. Also the
assumption of energy-independence of the tunneling matrix element does not necessarily hold true for
most practical junctions. A complete theory of  tunneling that takes into consideration all these special
features has been developed over the years \cite{simons}.
However, with the trick of absorbing the density of states into the tunneling resistance, the theory
described above is still useful, describing correctly the voltage dependence of the current. As we will
see below, the theory can be extended to the case of superconductors.
Similar tunneling theories can be constructed for scanning tunneling microscopy, where the densities of
state refer to the tip of the instrument and to the surface of the sample \cite{hofer}. Also, the
same structure has been successfully applied to systems such as trapped atoms, where the tunneling
can be realized by using RF fields coupled to internal hyperfine states of the atoms \cite{paivi}.
In this case, $\delta E$ is given by the detuning of the RF field with respect to the atomic frequency,
and the tunneling matrix element is  truly constant (given by the coupling amplitude of the field
with the atoms). In this situation, the current of particles between two hyperfine states depends indeed
on the volume of  the samples, as it should be, since the field penetrates the sample completely and
tunneling between the internal states occurs not only at some interface but in the whole volume. This
case also has its specifics however: the momentum (and not only the energy) is conserved.

In the case of superconductors, the theory proceeds as above, but the density of states used has to be replaced by that
given by the BCS theory.
It turns out that this density of states consists of a factor which is the same as for metals, which gets
multiplied by a divergent part near the gap.
It is then natural to introduce a normalized density of states ${\cal N}$, which would take the value 1 for metals,
\begin{eqnarray}
{\cal N}_{\rm normal}(\epsilon)
 &=& 1\,,
\end{eqnarray}
and will be a function
\begin{eqnarray}
{\cal N}_{\rm BCS}(\epsilon)
 &=& \frac{|\epsilon|}{\sqrt{\epsilon^2-\Delta^2}}\Theta(|\epsilon|-\Delta)\,,\label{eqn_BCS_DOS}
\end{eqnarray}
for BCS superconductors ($\Delta$ is the superconducting gap and $\Theta $ is the Heavyside function).

Summarizing all the discussion above, we can write the tunneling probability, including both the metallic
and superconducting case, as
\begin{subequations}\label{eqn_tunnel_prob}
\begin{eqnarray}
\Gamma_{1\rightarrow 2}(\delta \epsilon)
 &=& \frac{1}{e^2 R_{12}}\lint{-\infty}{\infty}{\epsilon_1}\hspace{-2mm}
                     \lint{-\infty}{\infty}{\epsilon_2}{\cal N}_1(\epsilon_1){\cal N}_2(\epsilon_2)f_1(\epsilon_1)[1-f_2(\epsilon_2)]
                     \delta(\epsilon_2-\epsilon_1-\delta E_{1\rightarrow 2})\label{eqn_tunnel_prob_a}\\
 &=& \frac{1}{e^2 R_{12}}\lint{-\infty}{\infty}{\epsilon_1}{\cal N}_{1}(\epsilon_1){\cal N}_2(\epsilon_1+\delta E_{1\rightarrow 2})f_1(\epsilon_1)[1-f_2(\epsilon_1+\delta E_{1\rightarrow 2})]\,,\label{eqn_tunnel_prob_b}
\end{eqnarray}
\end{subequations}
with ${\cal N}_{1,2}$ being the normalized density of states of the electrodes, as introduced above.
Furthermore, given the tunneling probabilities of Eq.\ (\ref{eqn_tunnel_prob}),
the current through a single superconducting tunnel junction without charging effects is given as
\begin{equation}
I_{1\rightarrow 2}
 = -e\left[\Gamma_{1\rightarrow2}(\delta E_{1\rightarrow 2})-\Gamma_{2\rightarrow1}(\delta E_{2\rightarrow 1})\right]
 = -e\left[\Gamma_{1\rightarrow2}(\delta E)-\Gamma_{2\rightarrow 1}(-\delta E_{1\rightarrow 2})\right]\,.
     \label{eqn_current_single_junction}
\end{equation}

\subsection{Numerical methods}
\label{num}

\begin{figure}[t]
\begin{center}
  \includegraphics[width=8cm]{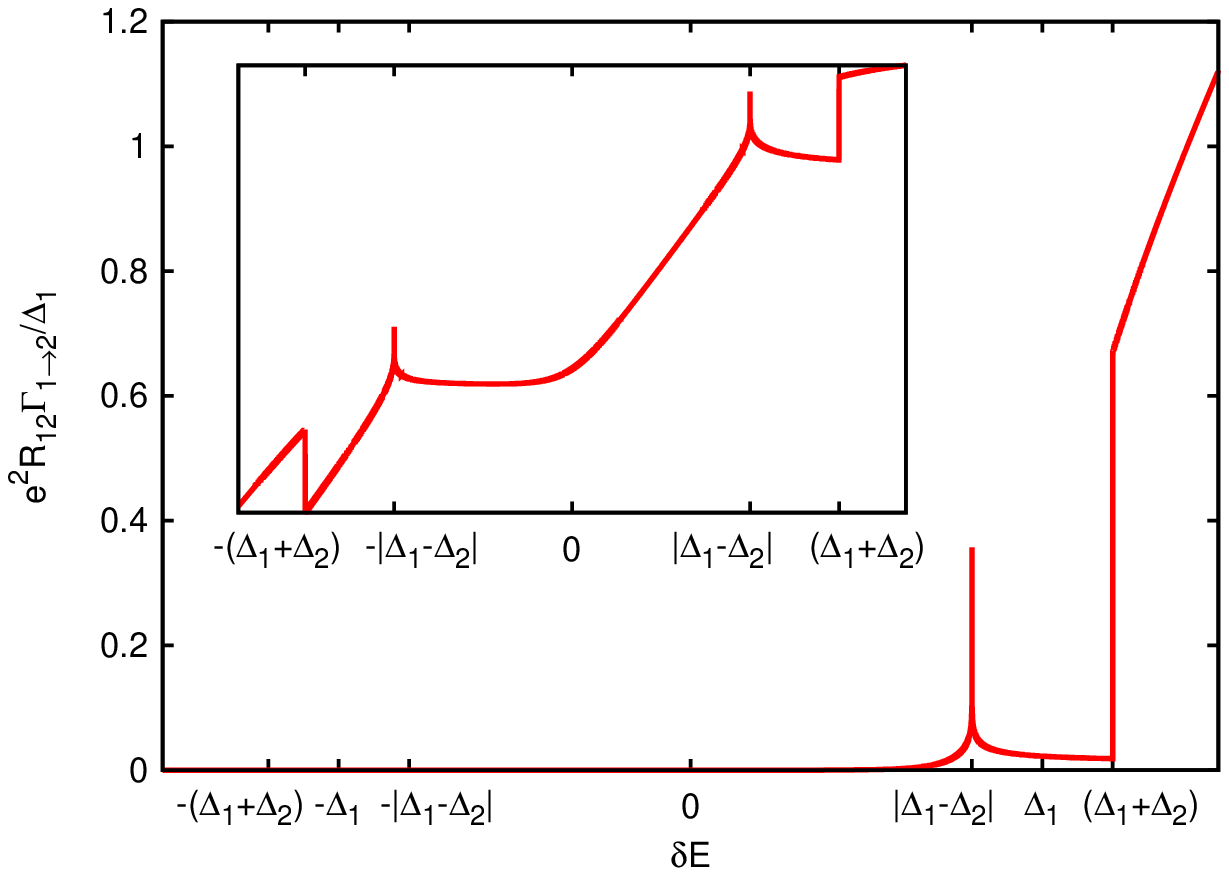}
  \includegraphics[width=8cm]{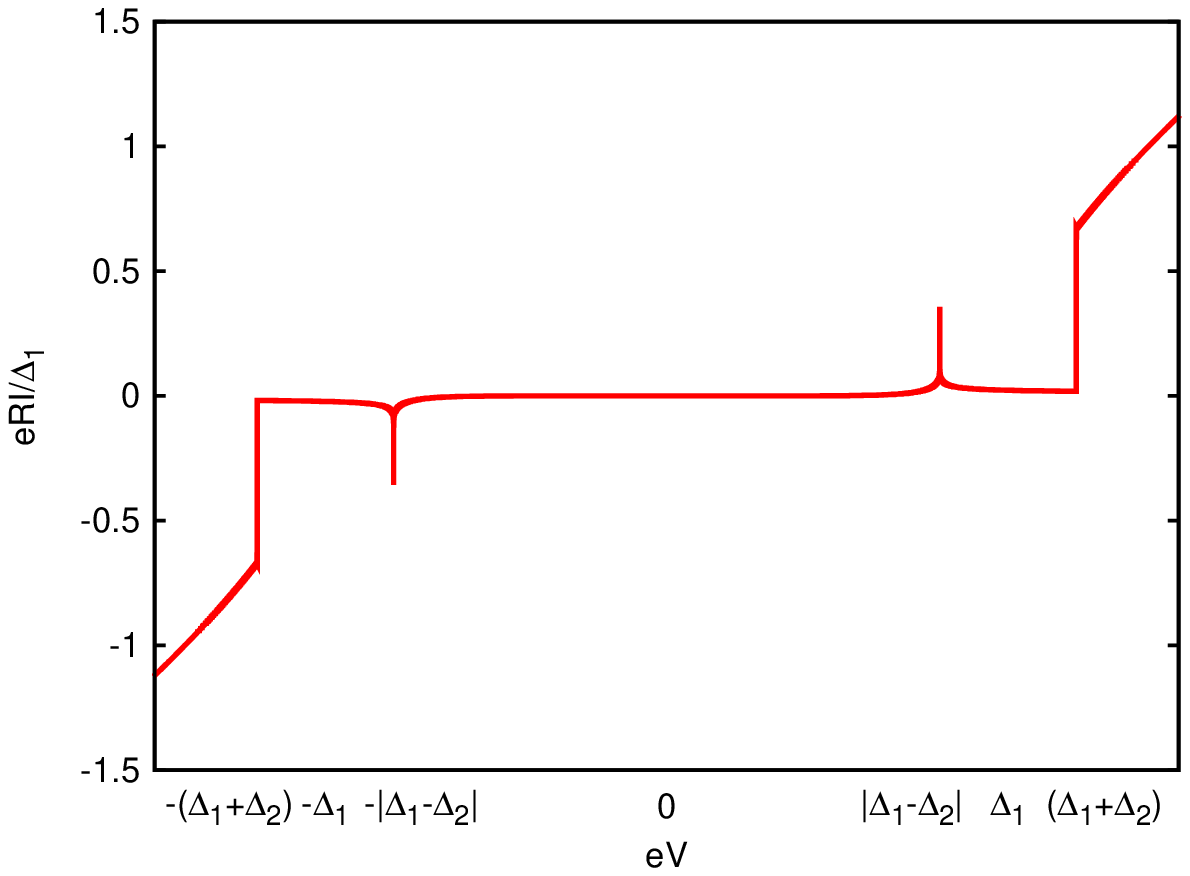}\\
  \includegraphics[width=8cm]{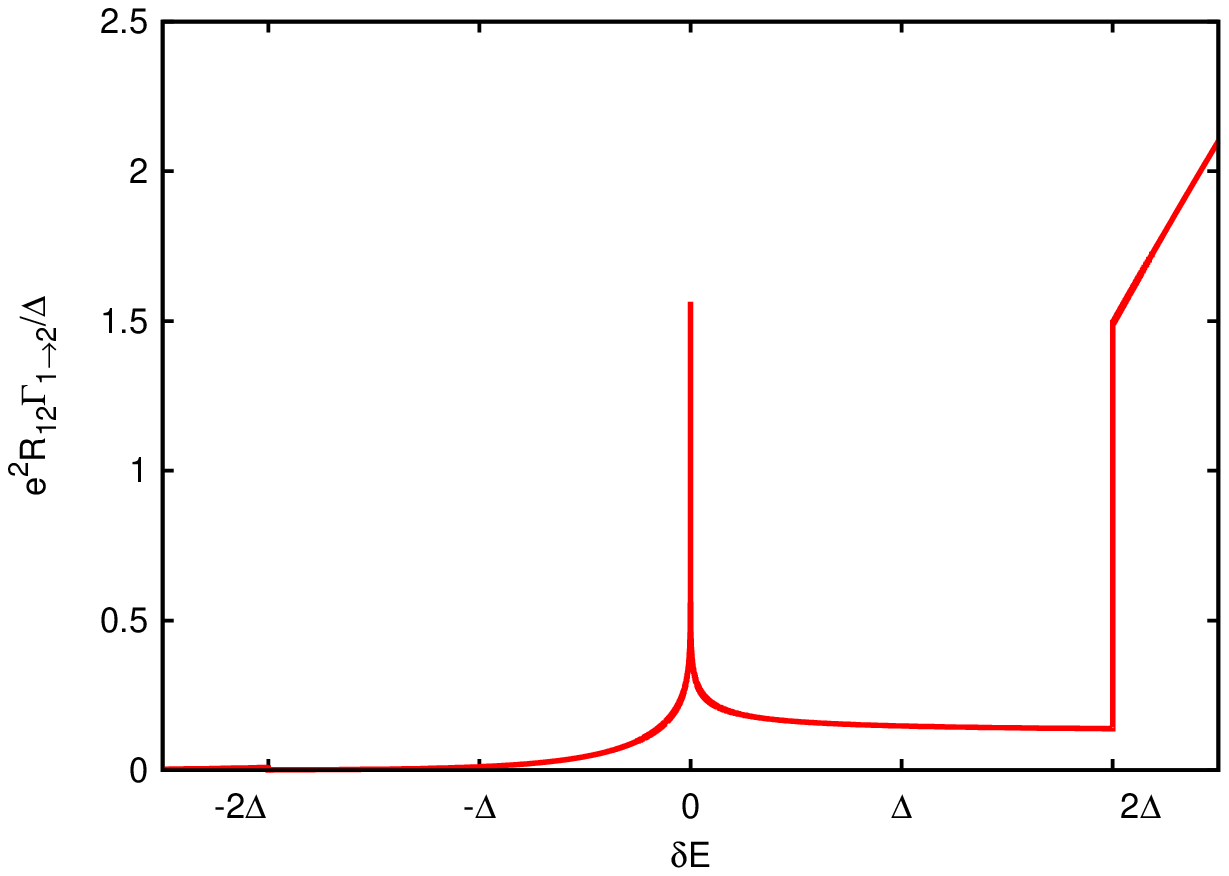}
  \includegraphics[width=8cm]{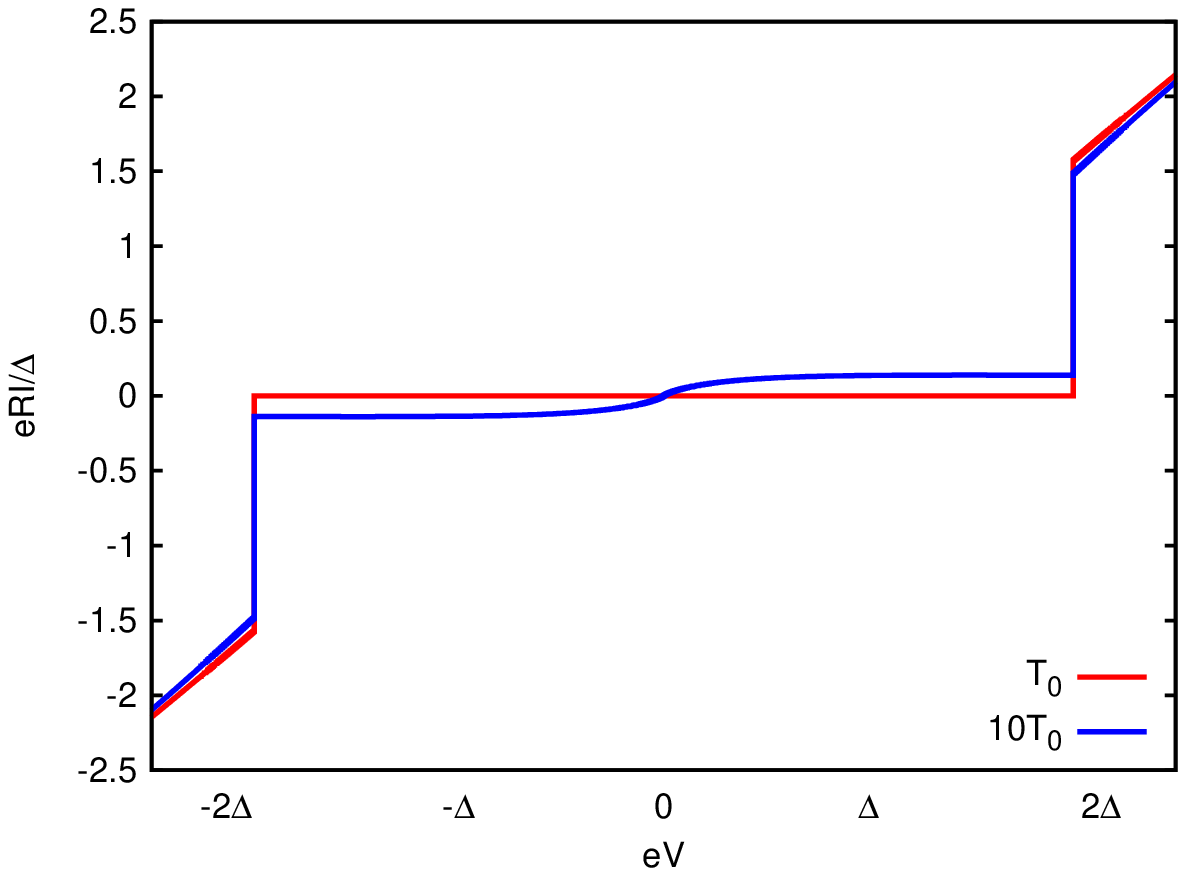}
\end{center}
\caption{Top: Tunneling rate (left) as a function of energy difference,
         and $I$--$V$ (right) through an asymmetric
         superconducting tunnel junction with $\Delta_2=0.2\Delta_1$.
         The inset in the tunnel rate plot shows the same
         data with a logarithmic scale on the y-axis, for better emphasis
         of all features.
         Bottom: Tunneling rate (left) and $I$--$V$ (right) through a symmetric superconducting tunnel
         junction. In the $I$--$V$ we show the current for two different
         temperatures, with $k_BT_0=3.92\times10^{-2}\Delta_1$.
         }\label{fig_gamma}
\end{figure}

For the discussion in this subsection we use the shorthand notation $\delta E = \delta E_{1\rightarrow 2}$.
Now Eq.\ (\ref{eqn_tunnel_prob}) can be solved analytically for some
limiting cases, as we have already seen
 for the case of normal metal electrodes, or can be expressed as a
combination of special functions. For example, in the case of a SIN junction, the calculation at zero temperature is
straightforward. From Eq. (\ref{eqn_tunnel_prob_b}) we get
\begin{equation}
\Gamma_{1\rightarrow 2} = \frac{1}{e^2 R_{12}}\sqrt{(\delta E)^2 - \Delta_{1}^2}\Theta (\delta E - \Delta_{1} ),
\end{equation}
which can be seen for example in Figure \ref{fig_gamma} bottom left (the additional spike appearing there at $\delta E = 0$
is a singularity-matching peak discussed below; it shows
up because the temperature is finite). In this figure, $I = I_{1 \rightarrow 2}$ and $\delta E = \delta E_{1\rightarrow 2} = -eV$.

In practice however, in the general case of non-identical superconductors
and finite temperature, it has to
be solved numerically, and, due to the singularities in the BCS density of
states, the
numerical treatment of Eq.\ (\ref{eqn_tunnel_prob}) is somewhat
challenging (see Fig.\ \ref{fig_gamma}).
We therefore want to discuss the numerical problems arising from these
divergencies and possible ways
to solve them in a bit more detail.
Inserting Eq.\ (\ref{eqn_BCS_DOS}) into Eq.\ (\ref{eqn_tunnel_prob_b}),
one can easily see that the
expression for the electron tunneling probability splits into four
separate integrals,
\begin{eqnarray}
\Gamma_{1\rightarrow 2}
 &=&
\frac{1}{e^{2}R_{12}}\left(\Upsilon_1+\Upsilon_2+\Upsilon_3+\Upsilon_4\right)\,,
\end{eqnarray}
where the integrals $\Upsilon_n$ are given by
\begin{subequations}\label{eqn_integral_split}
\begin{eqnarray}
\Upsilon_1
 &=& \lint{-\infty}{\min(-\Delta_1,-\Delta_2-\delta
E)\hspace{-20mm}}{\epsilon}\,g(\epsilon)\\
\Upsilon_2
 &=&-\Theta(-\Delta_1-\Delta_2 +\delta E)\lint{\Delta_2-\delta
E\hspace{-5mm}}{-\Delta_1}{\epsilon}g(\epsilon)\\
\Upsilon_3
 &=&-\Theta(-\Delta_1-\Delta_2-\delta E)\lint{\Delta_1}{-\Delta_2-\delta
E\hspace{-5mm}}{\epsilon}g(\epsilon)\\
\Upsilon_4
 &=& \lint{\max(\Delta_1,\Delta_2-\delta E)\hspace{-15mm}}{\infty}{\epsilon}g(\epsilon)\,.
\end{eqnarray}
\end{subequations}
The limits of these integrals are set by the $\Theta$-functions in the BCS
density of states. The function $g(\epsilon )$
is then
\begin{eqnarray}
g(\epsilon )
 &=& \frac{\epsilon (\epsilon +\delta E)}{\sqrt{\epsilon ^2-\Delta_1^2}\sqrt{(\epsilon +\delta
E)^2-\Delta_2^2}}f_1(\epsilon )[1-f_2(\epsilon +\delta E)]\,.
\end{eqnarray}
Due to the singularities in the BCS density of states, the function $g(\epsilon )$
is singular in each of the finite limits of the
integrals $\Upsilon_n$.

In most cases these singularities can be handled
easily enough with the help of a simple integral
transformation. As the function $g(\epsilon )$ behaves qualitatively like
$h(x)*x/\sqrt{x^2-c^2}$, where $h(x)$ is smooth on the
interval of integration, we can substitute $x=\sqrt{s^2+c^2}$ and we are left
with an integral which is numerically
easily handled:
\begin{equation}
\lint{c}{d}{x}\frac{xh(x)}{\sqrt{x^2-c^2}}=\lint{0}{\sqrt{d^2-c^2}}{s}h(\sqrt{s^2+c^2})\,.\label{eqn_int_trans}
\end{equation}
However, for certain values of $\delta E$, the singularities of the
densities of states of the first and second lead
will coincide, leading to two logarithmic divergences known as singularity
matching peaks at
$\delta E=\pm(\Delta_1+\Delta_2)$ and two finite steps at the
quasiparticle threshold
$\delta E=\pm|\Delta_1-\Delta_2|$.

\vspace{3mm}

{\it Singularity-matching peaks:} We will first discuss the singularity matching peaks, which are described by
the integrals $\Upsilon_1$ and $\Upsilon_4$.
When $\delta E$ is close to $\pm(\Delta_1+\Delta_2)$, the integration
results can become very inaccurate, and
the transformation done in Eq.\ (\ref{eqn_int_trans}) will not help to
solve the problem.
We can however separate the singularity from the
integral by partial integration, leaving us with a finite and easily
integrable rest term.
We note that the general form of integrals $\Upsilon_1$ and $\Upsilon_4$ is
\begin{eqnarray}
\Upsilon (c)
 &=& \lint{0}{\infty}{x}\frac{h(x)}{\sqrt{x(x+c)}}\,.
\end{eqnarray}
As $h(x)$ is proportional to the Fermi distribution, $h(x)$
and all
its derivatives exponentially tend to zero for large values of $x$.
Furthermore, $h(x)$ is analytic
on the interval $[0,\infty)$. Using partial integration, we obtain
\begin{eqnarray}
\Upsilon (c)
 &=& \ln\left(\frac{1}{c}\right)h(x=0)
    -2\lint{0}{\infty}{x}\ln\left(\sqrt{x}+\sqrt{x+c}\right)h^\prime(x)\,,\label{eqn_part_int}
\end{eqnarray}
with the big advantage that the singularity itself is described by an
explicit expression rather than an integral.
The integral in the second term is finite and easily computed. The major
nuisance here is to calculate the derivative
of the function $h(x)$.

For completeness, it is also possible to derive a complete expansion of
$\Upsilon (c)$ with the help of repeated
partial integration,
\begin{eqnarray}
\Upsilon (c)
 &=& \ln\left(\frac{1}{c}\right)h(x=0)
     -\sum_{n=1}^{\infty}\frac{(2n)!}{4^n(n!)^3}c^n\ln(c)h^{(n)}(x=0)\,.\label{eqn_expansion_of_Upsilon}
\end{eqnarray}
Here we denote by $h^{(n)}(x)$ the $n$th derivative of $h(x)$ with respect
to $x$ (for a derivation,
see Appendix \ref{sec_expansions}).

Using the recipe of Eq.\ (\ref{eqn_part_int}) on the integrals
$\Upsilon_{1}$ and $\Upsilon_{4}$, we obtain
\begin{subequations}\label{eqn_gammas_transformed}
\begin{eqnarray}
\Upsilon_1
 &=& S_1 -2 \hspace{-10mm}
\int\limits_{\textrm{max}(\Delta_1,\Delta_2+\delta
E)}^{\infty}\hspace{-10mm}d\epsilon
     \ln\left[\sqrt{\epsilon -\Delta_1}+\sqrt{\epsilon -\delta E-\Delta_2}\right]\nonumber\\
  && \hspace{25mm}\times\frac{d}{d\epsilon}\left[\frac{\epsilon (\epsilon -\delta E)}
     {\sqrt{(\epsilon +\Delta_1)(\epsilon -\delta E +\Delta_2)}}[1-f_1(\epsilon )]f_2(\epsilon -\delta
E)\right]\\
\Upsilon_4
 &=& S_4 -2 \hspace{-10mm}
\int\limits_{\textrm{max}(\Delta_1,\Delta_2-\delta
E)}^{\infty}\hspace{-10mm}d\epsilon
     \ln\left[\sqrt{\epsilon -\Delta_1}+\sqrt{\epsilon +\delta E-\Delta_2}\right]\nonumber\\
  && \hspace{25mm}\times\frac{d}{d\epsilon}\left[\frac{\epsilon (\epsilon+\delta E)}
     {\sqrt{(\epsilon +\Delta_1)(\epsilon +\delta E+\Delta_2)}}f_1(\epsilon )[1-f_2(\epsilon +\delta
E)]\right]\,,
\end{eqnarray}
\end{subequations}
where the terms $S_1$ and $S_4$ describe the singular points, with
different expressions if the singularity is approached
from above or below,
\begin{subequations}
\begin{eqnarray}
S_1(\delta E>\Delta_1-\Delta_2)
 &=& - \frac{\Delta_2(\Delta_2+\delta E)f_1(-\Delta_2-\delta E)f_2(\Delta_2)}
          {\sqrt{2\Delta_2(\Delta_1+\Delta_2+\delta E)}} \ln
(-\Delta_1+\Delta_2 +\delta E)\\
S_1(\delta E<\Delta_1-\Delta_2)
 &=& - \frac{\Delta_1(\Delta_1-\delta E)f_1(-\Delta_1)f_2(\Delta_1-\delta E)}
          {\sqrt{2\Delta_1(\Delta_1+\Delta_2-\delta E)}} \ln
(\Delta_1-\Delta_2-\delta E)\\
S_4(\delta E>\Delta_2-\Delta_1)
 &=&
     -\frac{\Delta_1(\Delta_1+\delta E)f_1(\Delta_1)f_2(-\Delta_1-\delta E )}
          {\sqrt{2\Delta_1(\Delta_2+\Delta_1+\delta E)}}\ln
(-\Delta_2+\Delta_1+\delta E )\\
S_4(\delta E<\Delta_2-\Delta_1)
 &=&
     -\frac{\Delta_2(\Delta_2-\delta E)f_1(\Delta_2-\delta E)f_2(-\Delta_2)}
          {\sqrt{2\Delta_2(\Delta_2+\Delta_1-\delta E)}}\ln
(\Delta_2-\Delta_1-\delta E)\,.
\end{eqnarray}
\end{subequations}
In the case of a symmetric junction, where $\Delta_1=\Delta_2=\Delta$,
only one singularity matching peak appears
in the tunneling rates (\ref{eqn_tunnel_prob}). In this case the
singularity matching peak disappears in
the current through the junction. With the derived expressions we can
easily understand this, as the expressions
$S_1$ and $S_4$ cancel out when inserted into the expression for the
current (\ref{eqn_current_single_junction}).
In this special case the current can therefore be computed using only the
respective second terms in
Eqs.\ (\ref{eqn_gammas_transformed}).

\vspace{3mm}

{\it Quasiparticle threshold:} The quasiparticle threshold is described by the integrals $\Upsilon_2$ and
$\Upsilon_3$ in
Eqs.\ (\ref{eqn_integral_split}). We can find an explicit expression for
the height of these
jumps with the help of a partial integration
and taking the limit $\delta E\rightarrow\pm(\Delta_1+\Delta_2)$.
Both $\Upsilon_2$ and $\Upsilon_3$ are of the general form
\begin{eqnarray}
\Upsilon (c) &=& \lint{-c}{c}{x}\frac{h(x)}{\sqrt{c^2-x^2}}\,,
\label{eqn_arctan_int}
\end{eqnarray}
where both $h(x)$ and $h^\prime(x)$ are analytic functions on the interval
$[-c,c]$. Then we can find
the solution of $\Upsilon (c)$ in the limit $c\rightarrow 0$ by
partial integration,
\begin{eqnarray}
\Upsilon_c
 &=& \frac{\pi}{2}\left[h(c)+h(-c)\right]
    -\lint{-c}{c}{x}\mathrm{arctan}\left[\frac{x}{\sqrt{c^2-x^2}}\right]h^\prime(x)\nonumber\\
 &\longrightarrow& \pi
h(0)+0\,,\hspace{5mm}\mathrm{for}\hspace{5mm}c\rightarrow0\,.
\end{eqnarray}
Using this method, we find that the height of the step at the quasiparticle threshold voltage  is
\begin{subequations}\label{eqn_finite_jumps}
\begin{eqnarray}
\frac{\pi}{2e^2R_{12}}\sqrt{\Delta_1\Delta_2}f_1(-\Delta_1)f_2(-\Delta_2)\,
\end{eqnarray}\label{eqn_finite_jumps_a}
for $\delta E=-(\Delta_1+\Delta_2)$, and
\begin{eqnarray}
-\frac{\pi}{2e^2R_{12}}\sqrt{\Delta_1\Delta_2}f_1(+\Delta_1)f_2(+\Delta_2)\,
\end{eqnarray}\label{eqn_finite_jumps_b}
\end{subequations}
for $\delta E=+(\Delta_1+\Delta_2)$.

We note here that at zero temperature the height of the first step is $(\pi/2e^{2}R_{12})\sqrt{\Delta_1\Delta_2}$ and the
second one is zero (due to the fact that Fermi functions at positive energies
are zero). The first peak is always more prominent and positive, while the second threshold is
observable only at finite temperatures and it appears as a negative step (see the inset of Figure \ref{fig_gamma}).

\vspace{3mm}

{\it Dynes parametrization:} As one can see from the above discussion, it is rather cumbersome to treat
superconducting tunnel junctions in the
BCS picture. The expressions for the tunneling probabilities of Eq.\
(\ref{eqn_tunnel_prob}) will also appear in
the treatment of superconducting SETs, where the strong features discussed
in this chapter will cause further
numerical difficulties, especially when computing conductances.
It is therefore convenient to somewhat smoothen the jumps and
singularities of the tunneling probabilities.

One method to achieve this is to replace the $\delta$-distribution in Eq.\
(\ref{eqn_tunnel_prob_a}) by
a Gaussian function of finite width. (This is not only a mathematical
trick but it corresponds, in the so-called $P(E)$ theory
\cite{ingold}
to the case of a high impedance environment at finite temperature).
One is then left to solve a two-dimensional integral with non-trivial integration limits.


A more convenient and at the same time physically more justifiable
solution to the problem is to introduce a broadening
parameter into the density of states (\ref{eqn_BCS_DOS}),
\begin{eqnarray}
{\cal N}_{\rm Dynes}
 &=&
\left|\Re\left\{\frac{E+i\eta}{\sqrt{(E+i\eta)^2+\Delta^2}}\right\}\right|\,.\label{eqn_dynes_DOS}
\end{eqnarray}
The parameter $\eta$ is called Dynes parameter \cite{dynes} and it
accounts for the finite life-time of quasiparticles in the
superconductor. The symbol $\Re$ stands for the real part of the
expression in curly brackets.
This broadening of the density of states leads to nonzero currents flowing
through the tunnel
junction at voltages smaller than $\Delta/e$ (say for a NIS junction), which are often seen in
experiments.
These sub-gap currents are believed to be due to additional states formed
inside the gap, allowing for the opening of new conduction channels. The
origin of these currents is still not clarified in  the literature. One should realize that
these subgap currents are distinct from the
currents due solely to the existence of a finite temperature (which causes
subgap excitations) in a superconductor.
Such  temperature effects are already included in the formulation
presented above.

It has been demonstrated \cite{ournb} that these
additional currents due to states formed inside  the gap are
Coulomb-blockaded as well (as it could be expected physically). Moreover,
they can be even used for Coulomb-blockade thermometry.
It should be noted also that an intrinsic dependence of the gap on
temperature and magnetic field exists, as given by standard BCS theory. By
analyzing the change in  the zero-bias conductance when either temperature
or magnetic  field is increased, it is possible to determine the critical
temperature and the critical magnetic field of the island \cite{jconf}.
A detailed description of another fitting procedure which uses the Dynes
form for the density of states is also given in {\it e.g.}\
\cite{koppinen_thermometry}.
Through replacing the density of states  in Eq.\ (\ref{eqn_BCS_DOS}) with
Eq.\ (\ref{eqn_dynes_DOS}) the tunneling probabilities
in Eq.\ (\ref{eqn_tunnel_prob}) become finite and continuous for all
values of $\delta E$ and integration takes
place on the whole real axis. The integral (\ref{eqn_tunnel_prob_b}) is
still not easy to solve with high precision,
but compared to the procedure that follows when using Eq.\
(\ref{eqn_BCS_DOS}), the numerical treatment becomes
certainly more simple. In the following sections we will exclusively use
Eq.\ (\ref{eqn_dynes_DOS}) as density of states, with a
small but finite value for $\eta$.

\section{Two-junction devices}
\label{trans}

Two-junction arrays are very useful devices for a variety of applications.
They are denoted usually as a heterostructure - for example, SINIS, NININ,
{\it etc.}, showing the phase of the electrons
(metallic or superconducting) at the operating temperature. If a gate is
capacitively coupled with the middle electrode (called island) by a coupling
capacitance $C_{g}$, then the device is usually referred to as a single electron transistor (SET).
If some or all of the leads in the SET are superconducting, the device will
have characteristics that stem altogether from
tunneling, charging effects, and superconducting gap.
The fabrication of such devices is possible due to  modern nanolithography
techniques, as mentioned in the previous section,
using which one can easily fabricate small junctions and small metallic or
superconducting grains, resulting in non-negligible charging energies.
For example, the area $S$ of  the junctions obtained by
shadow evaporation techniques can be of the order of 100 nm$^2$ or even less.
The
insulating layer is usually of the order of $d=$1 nm (usually 0.5-5 nm) in
thickness, and the dielectric constant $\epsilon_{r}$ of the oxide is of the
order of 10. Using the
formula for the capacitance $C=\epsilon_{0}\epsilon_{r}S/d$, we can readily obtain
capacitances of
the order of $C\approx 10^{-15}F$ (femtofarads). The energy scale
corresponding to this capacitance is $e^{2}/2C\approx 100$ $\mu V$ (or a
temperature of  the order of 1K).
Thus, charging effects can become observable at temperatures that can be
achieved by standard cryogenic devices, such as dilution refrigerators.
Effort has been put in recent times in reducing this capacitance even further,
so that charging effects would be visible at even higher
temperatures (preferably at room temperature \cite{tan}). This is where
novel fabrication methods, based on advances in nanoscience, are essential.

\subsection{Master equation}
\label{mmaster}

The Coulomb blockade phenomena discussed here are essentially classical. The
tunnel resistance is assumed to have a value much larger than the quanta of
resistance $R_{K}=h/e^{2}$. This makes the  number of electrons on an island
a well-defined classical quantity, albeit a discret one. The Coulomb
blockade effects then stem from the  fact  that the biasing  potential can
be varied continuosly  while the potential given by the product between the
charge and  the capacitance  can change only in steps. This gives rise to
steps in the current and Coulomb blockade oscillations in the conductance.
Except for tunneling, this phenomenon is indeed a classical one: the energy
level separation of a small metallic  island is still much smaller than the
thermal energy $k_{B}T$, therefore the spectrum can be treated as continuous. This should
be  contrasted to the case of tunneling in semiconductor nanostructures,
where the spectrum is discrete (see  \cite{beenakker}).

\begin{figure}[h]
  \begin{center}
    \begin{minipage}[t]{0.58\linewidth}
      \includegraphics[width=7.5cm]{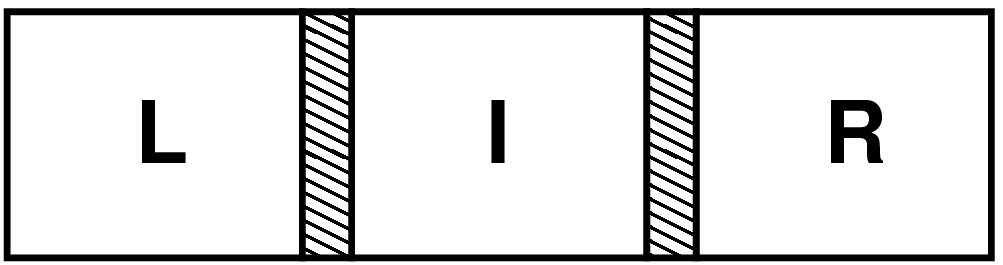}\vspace{5mm}\\
      \includegraphics[width=7.5cm]{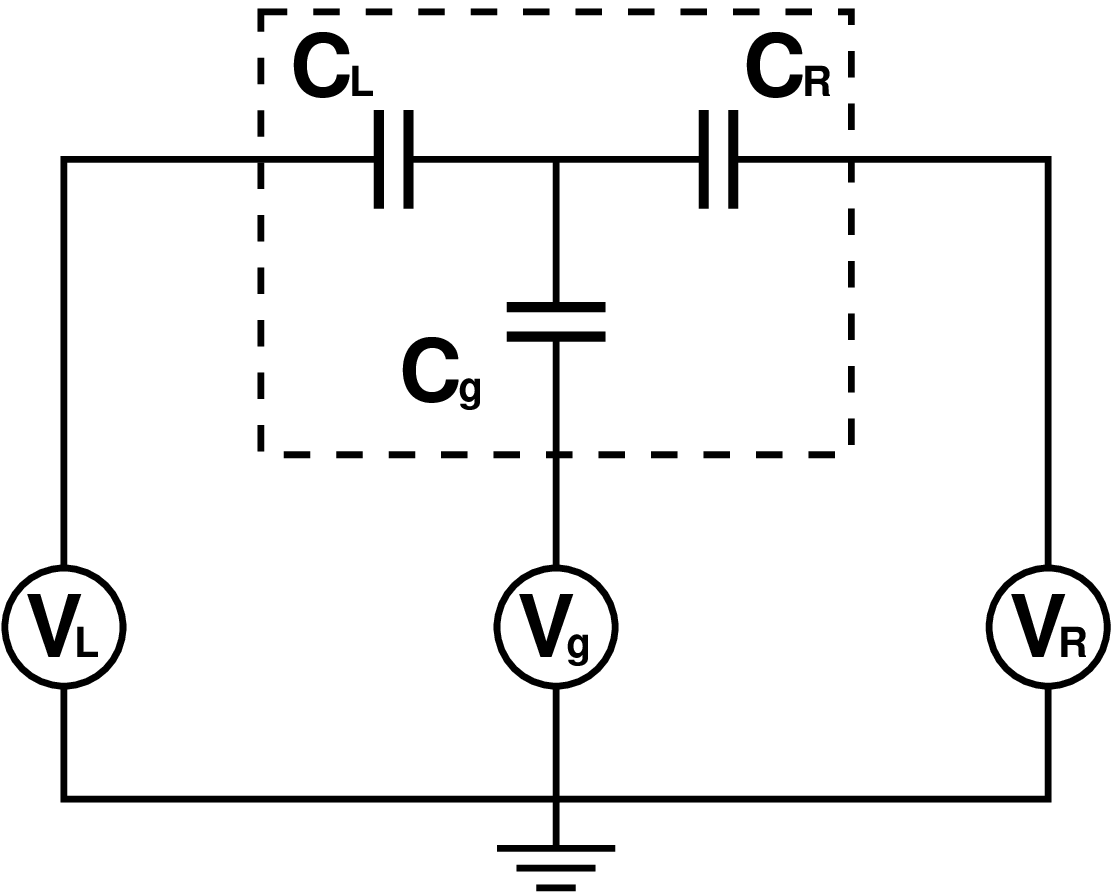}
    \end{minipage}\hfill
    \begin{minipage}[t]{0.38\linewidth}
      \caption{Schematic of an SET. The top picture shows the left (L) and right (R) leads
               in contact with the island (I) via a tunneling junction, each. The bottom picture shows the electrical
               equivalent circuit, where $V_{\rm L}$ and $V_{\rm R}$ are the voltages applied to the left and
               right lead, respectively, while $V_g$ is the gate voltage, that effectively alters
               the island charge.}\label{fig_set1}
    \end{minipage}
  \end{center}
\end{figure}
In the case of single electron transistors, the relevant charging energy is given by the effective capacitance
of the island to the ground: this is denoted in  the following, using an already standard notation, by $C_{\Sigma}$,
expressed as $C_\Sigma = C_L+C_R+C_g$. Here $C_L$ is the capacitance of the left junction, $C_{\rm R}$ is the capacitance
of the right junction, and $C_{g}$ is the capacitance between the  island and the gate electrode. The associated
charging energy of the island, corresponding to one electron of extra charge, is
\begin{eqnarray}
E_C
 &=& \frac{e^2}{2C_\Sigma}\,
\end{eqnarray}
The gate capacitance
$C_g$ is usually neglected or it can formally be distributed between the junction capacitances,
leaving $C_\Sigma \approx C_L+C_R$. The relevant electro-static energy of the island
with $n$ excess electrons and gated with the potential $V_{g}$  turns out to be
\begin{eqnarray*}
E_{\rm ch}(n,Q_g) &=&      \frac{1}{2C_\Sigma}\left(ne+q_g\right)^2\\
              &\equiv& E_C\left(n+n_g\right)^2\,,
\end{eqnarray*}
where $n_{g} \equiv q_{g}/e$ and $q_{g} \equiv C_{g}V_{g}+C_{\rm R}V_{\rm R}+C_{\rm L}V_{\rm L}$.
We want now to calculate the energy changes $\delta E_{\rm L\rightarrow I}(n)$, $\delta E_{\rm I\rightarrow L}(n)$, $\delta E_{\rm R\rightarrow I}(n)$, and $\delta E_{\rm I\rightarrow R}(n)$,
(including the work done by the sources) associated with electrons tunneling onto and off the
island). The general idea is that, when such processes occur, $n$ is increased (or decreased), respectively, by one, and the energy difference between the electron states before and after the tunneling event will be given
by $E_{ch}(n+1)-E_{ch}(n)$ (or $E_{ch}(n)-E_{ch}(n-1)$) plus the work done by the voltage source.

We give now a detailed derivation of these results. Kirchhoff's laws for the circuit presented in
Eq.(\ref{fig_set1}) can be written as:
\begin{subequations}\label{eqn_charge_balance}
\begin{eqnarray}
-ne &=& Q_{\rm L} + Q_{\rm R} + Q_{g}, \label{11}\\
\frac{Q_{\rm L}}{C_{\rm L}} - \frac{Q_{\rm R}}{C_{\rm R}} &=& V_{\rm L}-V_{\rm R}, \label{22}\\
\frac{Q_{\rm L}}{C_{\rm L}} - \frac{Q_{g}}{C_{g}} &=& V_{\rm L}-V_{g}, \label{33}\\
\frac{Q_{\rm R}}{C_{\rm R}} - \frac{Q_{g}}{C_{g}} &=& V_{\rm R}-V_{g}, \label{44}
\end{eqnarray}
\end{subequations}
where $Q_{\rm L}$, $Q_{\rm R}$ and $Q_{g}$ are the charges corresponding to capacitors $C_{\rm L}$, $C_{\rm R}$ and $C_{g}$ respectively.
Combining these equations, we find
\begin{subequations}\label{eqn_charge_distribution}
\begin{eqnarray}
Q_{g}(n) &=& -\frac{C_{g}}{C_{\Sigma}}\left[C_{\rm R}V_{\rm R} + C_{\rm L}V_{\rm L} - V_{g}(C_{\rm L}+C_{\rm R}) + ne\right] \label{am}\\
Q_{\rm L}(n) &=& -\frac{C_{\rm L}}{C_{\Sigma}}\left[C_{\rm R}V_{\rm R} + C_{g}V_{g} - V_{\rm L}(C_{g}+C_{\rm R})+ne\right] \label{bm}\\
Q_{\rm R}(n) &=& -\frac{C_{\rm R}}{C_{\Sigma}}\left[C_{\rm L}V_{\rm L} + C_{g}V_{g} -V_{\rm R}(C_{g}+C_{\rm L}) +ne\right]\label{cm}
\end{eqnarray}
\end{subequations}
The total electrostatic energy of the system stored in the capacitors is given by
\begin{equation}
E^{\rm (el)} (n)= \frac{Q_{\rm L}^{2}}{2C_{\rm L}} + \frac{Q_{\rm R}^{2}}{2C_{\rm R}} + \frac{Q_{g}^{2}}{2C_{g}} \label{elec}
\end{equation}
Using the expressions  Eqs.\ (\ref{eqn_charge_distribution}) one can verify immediately
an interesting property of $E^{(el)} (n)$, namely that it does not contain any terms linear in $n$,
\begin{equation}
E^{\rm (el)} (n) = \frac{(ne)^2}{2C_{\Sigma}} + {\rm terms \ depending \ only \ on \ voltages}.
\end{equation}

Consider now a process by which one electron tunnels from the left electrode to the island, accompanied by
a re-arrangement of the charges so that electrostatic equilibrium is reached corresponding to a state with
$n+1$ extra electrons on the island. The work done by the sources is then given by
\begin{equation}
W_{\rm L\rightarrow I} = \left[Q_{\rm L}(n+1)-Q_{\rm L}(n) -e\right]V_{\rm L}+V_{\rm R}\left[Q_{\rm R}(n+1)-Q_{\rm R}(n)\right]
+V_{g}\left[Q_{g}(n+1)-Q_{g}(n)\right].
\end{equation}
Let us look a bit in slow motion  at what is in fact happening here: as the electron tunnels through the left
island, to ensure the neutrality of the conductor connecting the source $V_{\rm L}$ to the left island, another
electron must be pulled through the source, which requires the energy $-eV_{\rm L}$. As the electron arrives
on the island, the system is no longer in the electrostatic equilibrium ensured by Eqs.\ (\ref{eqn_charge_balance})
and, to reach the new equilibrium with $n+1$ electrons on the island, the charges $Q_{\rm L}(n+1)-Q_{\rm L}(n)$,
$Q_{\rm R}(n+1)-Q_{\rm R}(n)$, and $Q_{g}(n+1)-Q_{g}(n)$ are being transferred through the corresponding sources.
The total energy change during this process, which includes the work done by the sources,
is
\begin{equation}
\delta E_{\rm L\rightarrow I} = W_{\rm L\rightarrow I} + E^{\rm (el)}(n)-E^{\rm (el)}(n+1).
\end{equation}
In other words, when an electron tunnels from a state $k_{\rm L}$ of the left electrode to a state $q_{I}$
of the island, the energy is conserved (due to the Dirac delta appearing in the transition rate) in  the
following way: $\epsilon_{q_{I}} = \epsilon_{k_{\rm L}} +\delta E_{\rm L\rightarrow I}$, or
$\epsilon_{q_{I}} + E^{\rm (el)}(n+1)= \epsilon_{k_{\rm L}}+ W_{\rm L\rightarrow I} + E^{\rm (el)}(n)$.
There are altogether four different possible tunneling events, two increasing and two decreasing
the island charge:
\begin{subequations}\label{eqn_tunnel_energies}
\begin{eqnarray}
    \delta E_{\rm L\rightarrow I}(n)  &=& -eV_L - 2E_C\left(n+n_g + 1/2\right)\,, \label{at}\\
    \delta E_{\rm I\rightarrow L}(n)  &=& +eV_L + 2E_C\left(n+n_g - 1/2\right)\,, \label{bt}\\
    \delta E_{\rm I\rightarrow R}(n)  &=& +eV_R + 2E_C\left(n+n_g - 1/2\right)\,, \label{ct}\\
    \delta E_{\rm R\rightarrow I}(n)  &=& -eV_R - 2E_C\left(n+n_g + 1/2\right)\,. \label{dt}
\end{eqnarray}
\end{subequations}
Another handy way to get these changes in energy is to introduce the free energy of the system as
the Legendre transform of the electrostatic energy Eq. (\ref{elec})
\begin{equation}
E_{\rm ch}(n) = E^{\rm (el)}(n) - Q_{\rm L}(n)V_{\rm L} - Q_{\rm R}(n)V_{\rm R} - Q_{g}(n)V_{g} .
\end{equation}
Then, using Eqs. (\ref{eqn_charge_distribution}) we get (up to a constant)
\begin{equation}
E_{\rm ch}(n) = \frac{1}{2C_{\Sigma}}\left(ne + q_{g}\right)^2 ,
\end{equation}
and the expressions Eqs. (\ref{eqn_tunnel_energies}) can be put in a physically more transparent form
\begin{subequations}\label{eqn_tunnel_energies_direct}
\begin{eqnarray}
    \delta E_{\rm L\rightarrow I}(n)  &=& -eV_L + E_{\rm ch}(n) - E_{\rm ch}(n+1)\, ,\\
    \delta E_{\rm I\rightarrow L}(n)  &=& +eV_L + E_{\rm ch}(n) - E_{\rm ch}(n-1)\, ,\\
    \delta E_{\rm I\rightarrow R}(n)  &=& +eV_R + E_{\rm ch}(n) - E_{\rm ch}(n-1)\, ,\\
    \delta E_{\rm R\rightarrow I}(n)  &=& -eV_R + E_{\rm ch}(n) - E_{\rm ch}(n+1)\, .
\end{eqnarray}
\end{subequations}

We see that $E_{\rm ch}(n)$ has the meaning of  an  effective charging energy of the island,
including the work done by the sources.

The probability to tunnel onto or off the island through either the left or right junction can be obtained
by inserting the respective energy difference of Eq.\ (\ref{eqn_tunnel_energies}) into Eq.\ (\ref{eqn_tunnel_prob}).
We define the probabilities for the four different tunneling events as function of the excess charge $n$ on the
island as
\begin{subequations}\label{eqn_set_gammas}
\begin{eqnarray}
    \Gamma_{\rm L\rightarrow I}(n) &\equiv& \Gamma[\delta E_{\rm L\rightarrow I}(n)]\\
    \Gamma_{\rm I\rightarrow L}(n) &\equiv& \Gamma[\delta E_{\rm I\rightarrow L}(n)]\\
    \Gamma_{\rm I\rightarrow R}(n) &\equiv& \Gamma[\delta E_{\rm I\rightarrow R}(n)]\\
    \Gamma_{\rm R\rightarrow I}(n) &\equiv& \Gamma[\delta E_{\rm R\rightarrow I}(n)]\,.
\end{eqnarray}
\end{subequations}

To continue our analysis, we now notice that, due to the fact that the number of electrons on the
island is a well-defined number, we can talk about classical states of the system and index them with the
excess number of  electrons on the island, $n$. There will then be transitions between these states,
namely, with the notations in  Eqs.\ (\ref{eqn_set_gammas}), the rates corresponding to the
the island's charge number $n$ being raised or lowered by one ($\Gamma_{\uparrow}$
and $\Gamma_{\downarrow}$ respectively) become,
\begin{subequations}\label{eqn_gamma_up_down}
\begin{eqnarray}
    \Gamma_\downarrow(n) &=& \Gamma_{I\rightarrow L}(n)+\Gamma_{I\rightarrow R}(n)\label{ohoo}\\
    \Gamma_\uparrow(n) &=& \Gamma_{\rm L\rightarrow I}(n)+\Gamma_{\rm R\rightarrow I}(n)\,.\label{ohooo}
\end{eqnarray}
\end{subequations}
We now want to calculate the rate at which the probability $p(n;t)$ that the system is in state $n$
changes. There are three contributions to this rate:
\begin{itemize}
\item If the system is already in the state $n$, it can
make transitions to the state $n+1$ or to the state $n-1$, with transition rates $\Gamma_\uparrow(n)$ and
$\Gamma_\downarrow(n)$;
\item The system can get to the state $n$ from state $n-1$ with transition rate
$\Gamma_\uparrow(n-1)$;
\item The system can get to the state $n$ from state $n+1$ with transition rate
$\Gamma_\downarrow(n+1)$.
\end{itemize}
This results in a master equation for $p(n;t)$,
\begin{equation}
\frac{dp(n;t)}{dt} = - [\Gamma_{\uparrow}(n) + \Gamma_{\downarrow}(n)]p(n;t) +  \Gamma_{\downarrow}(n+1)p(n+1;t)
+\Gamma_{\uparrow}(n-1)p(n-1;t) . \label{master}
\end{equation}
This equation describes  a classical Markovian process: indeed, embedded  in this treatment is the idea that
electrons do not have any memory while tunneling. The history of the electrons being at previous times in other
states is erased and therefore
there is no time-dependence in the transition rates Eqs.\ (\ref{eqn_gamma_up_down}), which depend only on the
"present" state $n$.

We now want to characterize the stationary state, defined by the condition $d p(n;t)/dt = 0$. From
Eq. (\ref{master}) we can verify immediately that the stationary state is given by a time-independent
$p(n)$ satisfying
\begin{subequations}\label{eqn_probabilities_n}
\begin{equation}
p(n)\Gamma_\uparrow(n)=p(n+1)\Gamma_\downarrow(n+1), \label{st1}
\end{equation}
which is equivalent to
\begin{equation}
p(n)\Gamma_\downarrow(n)=p(n-1)\Gamma_\uparrow(n-1), \label {st2}
\end{equation}
\end{subequations}
Using now the normalization condition for probabilities,
\begin{equation}
   \sum\limits_{n=-\infty}^{\infty}p(n) = 1,
  \label{eqn_pn_master}
\end{equation}
we find
\begin{eqnarray}
p(n) &=& \frac{\left[\prod\limits_{i=-\infty}^{n-1}\Gamma_\uparrow(i)\right]
	       \left[\prod\limits_{i=n+1}^{\infty}\Gamma_\downarrow(i)\right]}
	      {\sum\limits_{m=-\infty}^{\infty}\left(
	       \left[\prod\limits_{i=-\infty}^{m-1}\Gamma_\uparrow(i)\right]
	       \left[\prod\limits_{i=m+1}^{\infty}\Gamma_\downarrow(i)\right]
               \right)}.\label{eqn_pn_direct}
\end{eqnarray}
From Eqs. (\ref{eqn_probabilities_n}) another useful property results immediately,
\begin{equation}
\sum\limits_{n=-\infty}^{\infty} p(n)\Gamma_{\uparrow}(n) = \sum\limits_{n=-\infty}^{\infty} p(n)\Gamma_{\downarrow}(n) \label{qqq}
\end{equation}
Using now  Eq. (\ref{qqq}) and the definitions in Eqs. (\ref{eqn_gamma_up_down}) it is easy to check  that there
is no charge accumulation on the island,
that is, the currents flowing through the left junction and the right junction are equal,
\begin{equation}
I=I_{\rm L\rightarrow I}=I_{I\rightarrow L} = - e\sum\limits_{n=-\infty}^{\infty}p(n)\left[\Gamma_{\rm L\rightarrow I}(n)
-\Gamma_{I\rightarrow L}(n)\right] = - e\sum\limits_{n=-\infty}^{\infty}p(n)\left[\Gamma_{I\rightarrow R}(n)
-\Gamma_{\rm R\rightarrow I}(n)\right]
\label{eqn_current}
\end{equation}
This formula is used in the following to calculate the currents and the conductivities.

We point out here that the simplifications leading to  Eqs. (\ref{eqn_pn_master}-\ref{eqn_current})  are valid only under the assumption of steady-state
conditions. In cases in which biasing is done by RF fields -  such as turnstile operation or RF cooling \cite{kafanov} - this assumption could break and the treatment above is insufficient.

\subsection{Coulomb blockade thermometry}
\label{cbl}
\begin{figure}[ht]
\begin{center}
  \includegraphics[width=8cm]{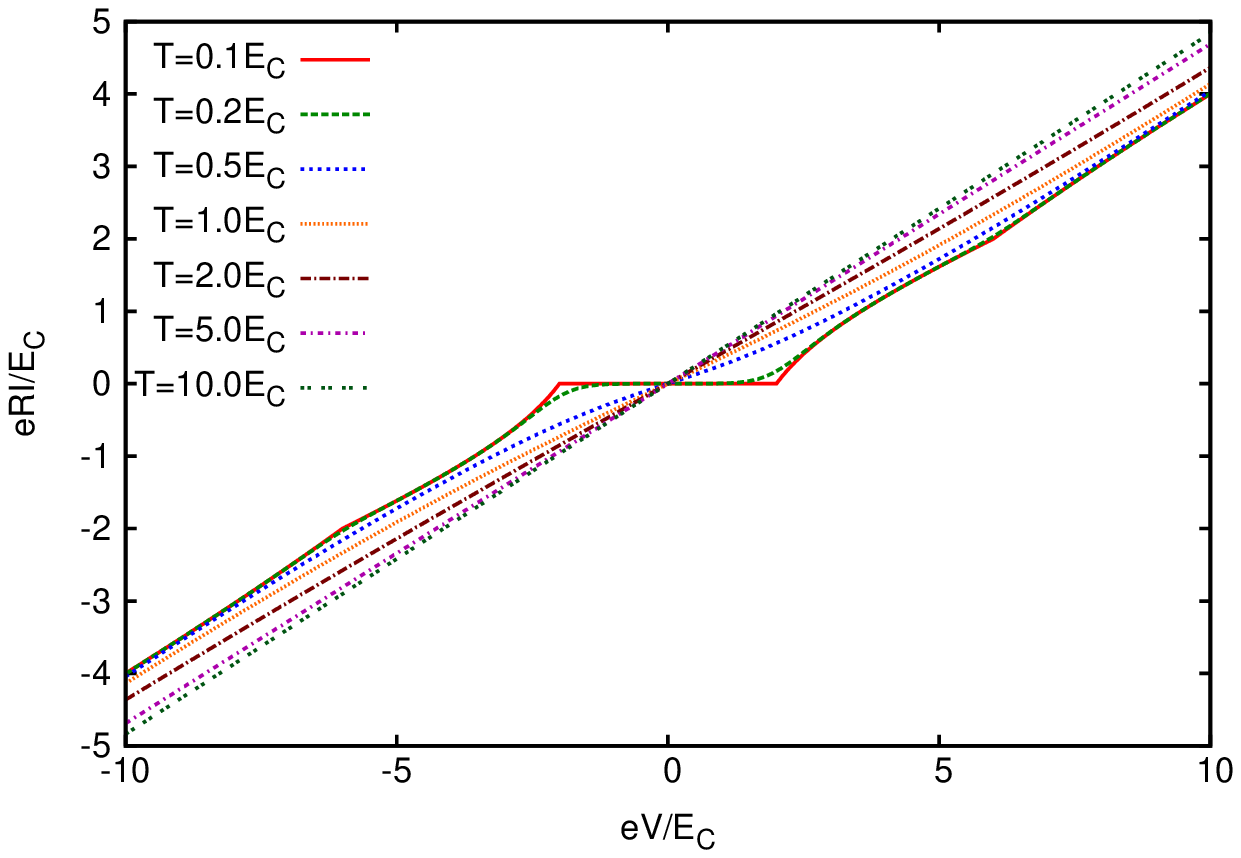}
  \includegraphics[width=8cm]{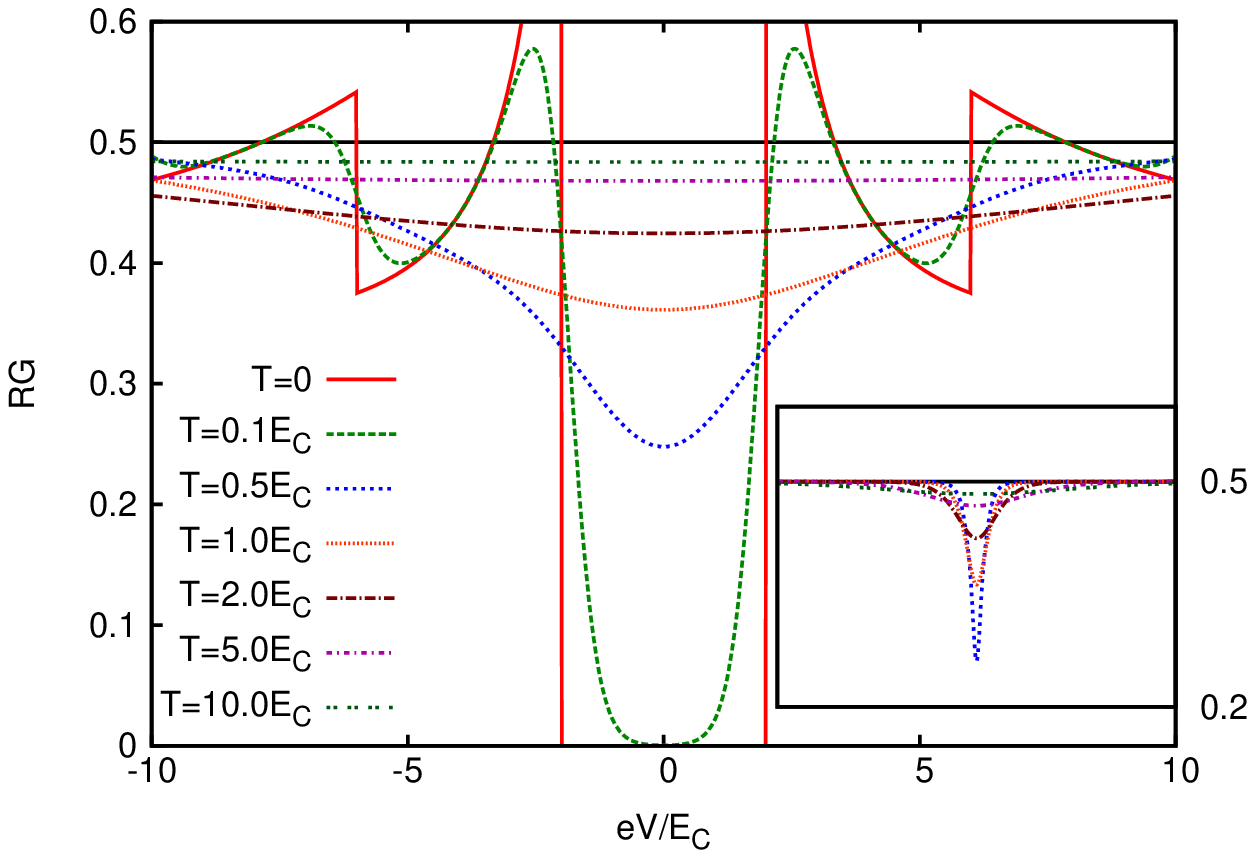}
\end{center}
\caption{The current (left) and the conductance (right) of a Coulomb blockade thermometer at
different temperatures. For temperatures $T\leq E_C$ one can
clearly see the steps in the current and the oscillations in the conductance. For $T\geq
E_C$ these steps/osillations smear out and the Coulomb dip forms. The inset in the
conductance plot shows the Coulomb dip curves ($T\geq E_C$) in a voltage range of
$[-100E_C,100E_C]$ to make the typical shape of the dip also visible for higher
temperatures.}
\label{fig_cbt}
\end{figure}

A very useful application of double junction structures with normal-metal electrodes (NININ)
is for temperature measurements. Consider such a structure, with charging energy of the island
denoted as before by $E_{C} = e^2 /2C_{\Sigma}$ and at temperature (assuming equilibrium) $T$. Clear Coulomb-blockade
effects can be seen if  the temperature is much smaller than the charging energy: the current is zero for voltages
lower then $e/C_{\Sigma}$, as it takes that much energy to add an electron on the island. Plateaus are then formed
at values of the bias voltage corresponding to multiples of $e/C$ (see
Figure \ref{fig_cbt} left).
The conductance curves then show so-called Coulomb-blockade oscillations (see Figure \ref{fig_cbt}) right). Gate
voltages can be also included in this analysis, in which case it can be shown that, in the gate voltage -- bias
voltage space the stable states (corresponding to different numbers of electrons on the island) form a diamond
pattern \cite{al}.

We now examine the effect of temperature: intuitively, as the temperature is increased, one expects
that the features visible at low temperatures will be smeared out and eventually washed away. Surprisingly
however, if the temperature is larger than the charging energy, one clear feature
survives in the conductance, namely a dip at low bias voltages. It can be shown \cite{cbt}, within the
orthodox theory presented above, that, in the limit $k_{B}T\gg E_C$, the full width of this
Coulomb blockade dip  (measured half-way between the dip in the conductance at zero bias and the plateau
at $V \rightarrow \pm \infty$)
is given by
\begin{equation}
\frac{eV_{1/2}}{2k_{B}T} \approx 5.439 , \label{prim}
\end{equation}
and the relative change of conductance at zero bias voltage is
\begin{equation}
\frac{G(V\rightarrow \pm\infty ) - G(V=0)}{G (V\rightarrow \pm\infty )} = \frac{E_{C}}{3K_{B}T}.\label{prs}
\end{equation}
It is straigthforward to show that $G(V\rightarrow \pm\infty ) = (2 R)^{-1}$, meaning that at large bias voltage
the charging energy of the island does not have any effect and the conductance of the NININ array becomes that
of two junctions of resistance $R$ in series. Eq. (\ref{prim}) enables the use of such structures as primary thermometers: from a simple measurement
of conductance it is possible to extract $V_{1/2}$ without any need for calibration. Eq. (\ref{prs}) can be used to determine $E_{C}$.
\begin{figure}[h]
\begin{center}
\includegraphics[width=9cm]{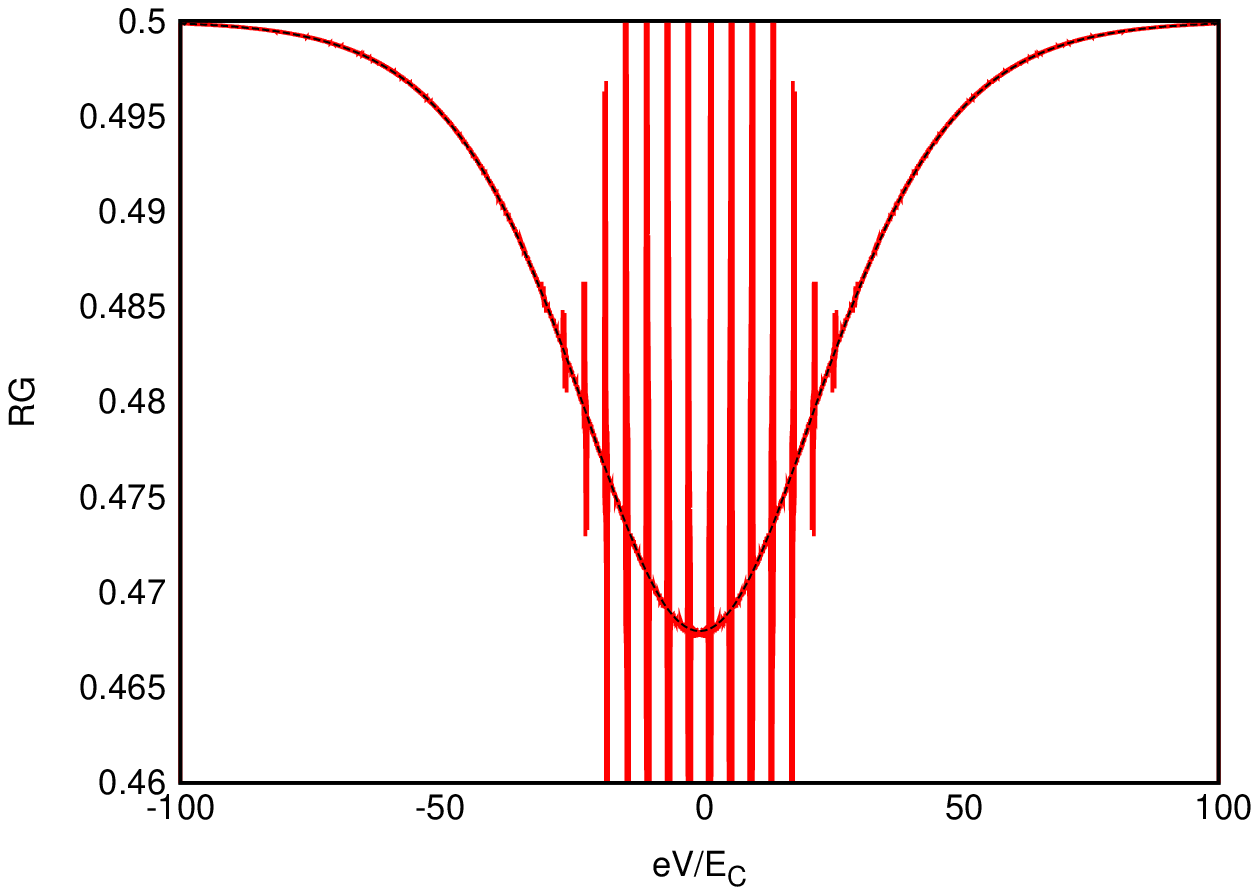}
\includegraphics[width=8cm]{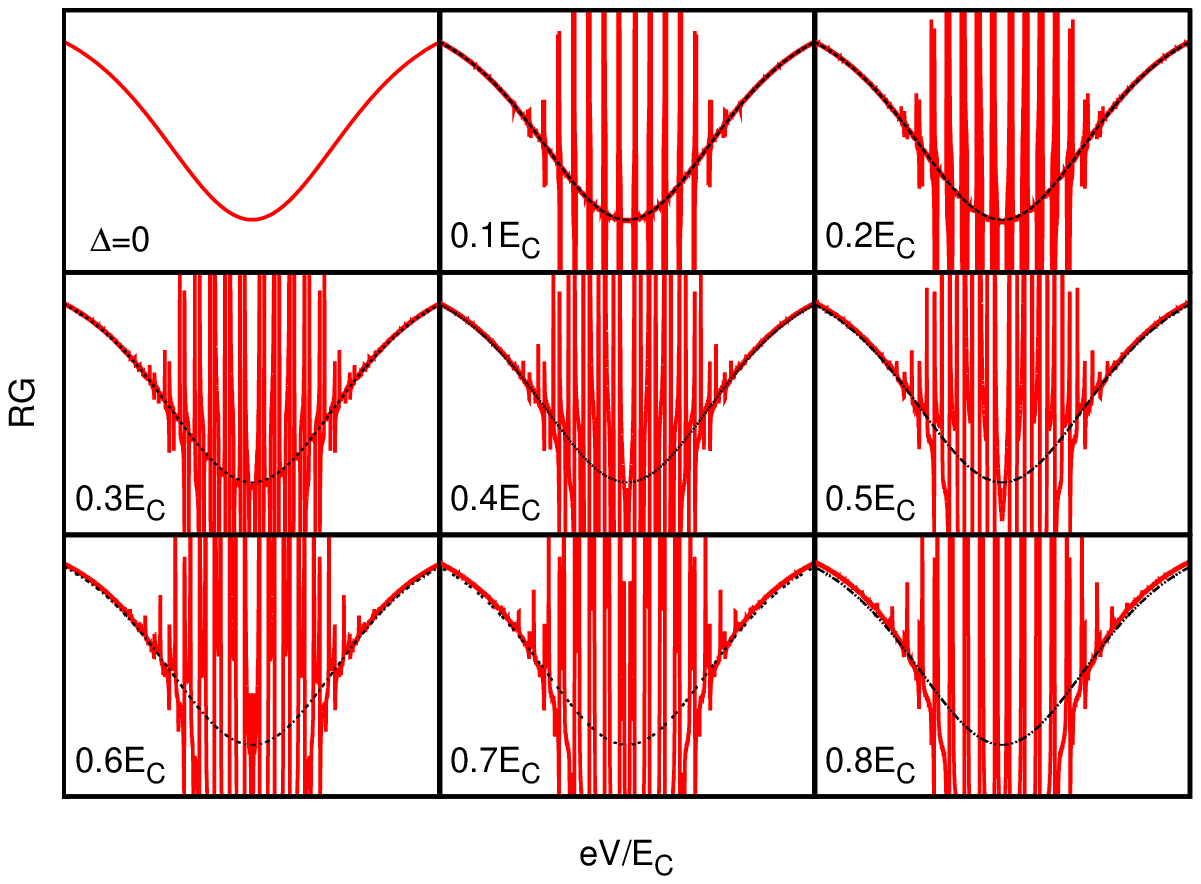}
 \end{center}
\caption{Left: The conductance dip in a superconducting CBT
at a temperature of $T=5E_C$, with a finite gap
$\Delta_I=\Delta_L=0.1E_C$ (red color). For
comparison we overlayed the graph with the corresponding curve for the all-metallic case (black line).  The CB dip is still clearly visible, but the features due to superconductivity are clearly
imprinted on the curve, according to Eq.\ (\ref{eqn_peaks}).
Right: The conductance dip in a superconducting CBT,
with a growing superconducting gap from $0$ to $0.8E_C$. Here we can see
how the dip is gradually smeared out by the features that arise due to the
superconducting nature of the island and leads.}
\label{uuup}
\end{figure}

What happens now if  the electrodes become superconducting?
In Figure \ref{uuup} left we present the behavior of the Coulomb blockade dip
for a SINIS structure as the gap of the electrodes is increased. One can see how the superconducting gap is
introducing additional features into the conductance curve which, as the gap is gradually increased, finally
distort its shape so much that the CBT dip is very soon not distinguishable anymore. The origin  of these features is described  in detail in Section \ref{sset}.

\subsection{SINIS}
\label{sinis}
A superconductor-insulator-normal metal-insulator-superconducting (SINIS) device can be seen as an SET without gate,
with superconducting leads and a normal metal island. SINIS are very useful devices, used as local coolers
\cite{microcoolers}, secondary thermometers \cite{koppinen_thermometry}, and electron pumps \cite{pump}.
Usually the charging energy in a SINIS is so small that it can be neglected. However, in some applications,
for instance as thermometers, the
SINIS dimensions may become small enough to make charging effects become observable.
\begin{figure}[t]
\begin{center}
  \includegraphics[width=8cm]{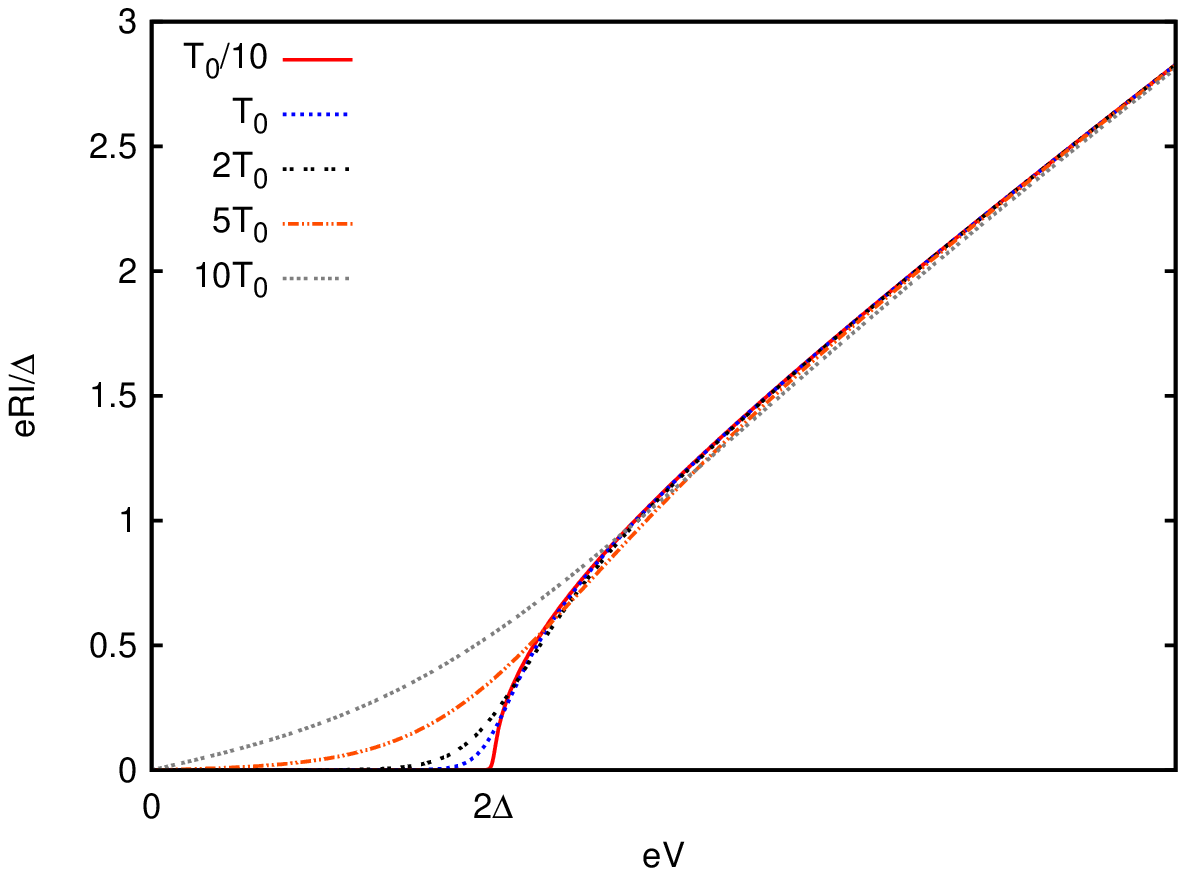}
  \includegraphics[width=8cm]{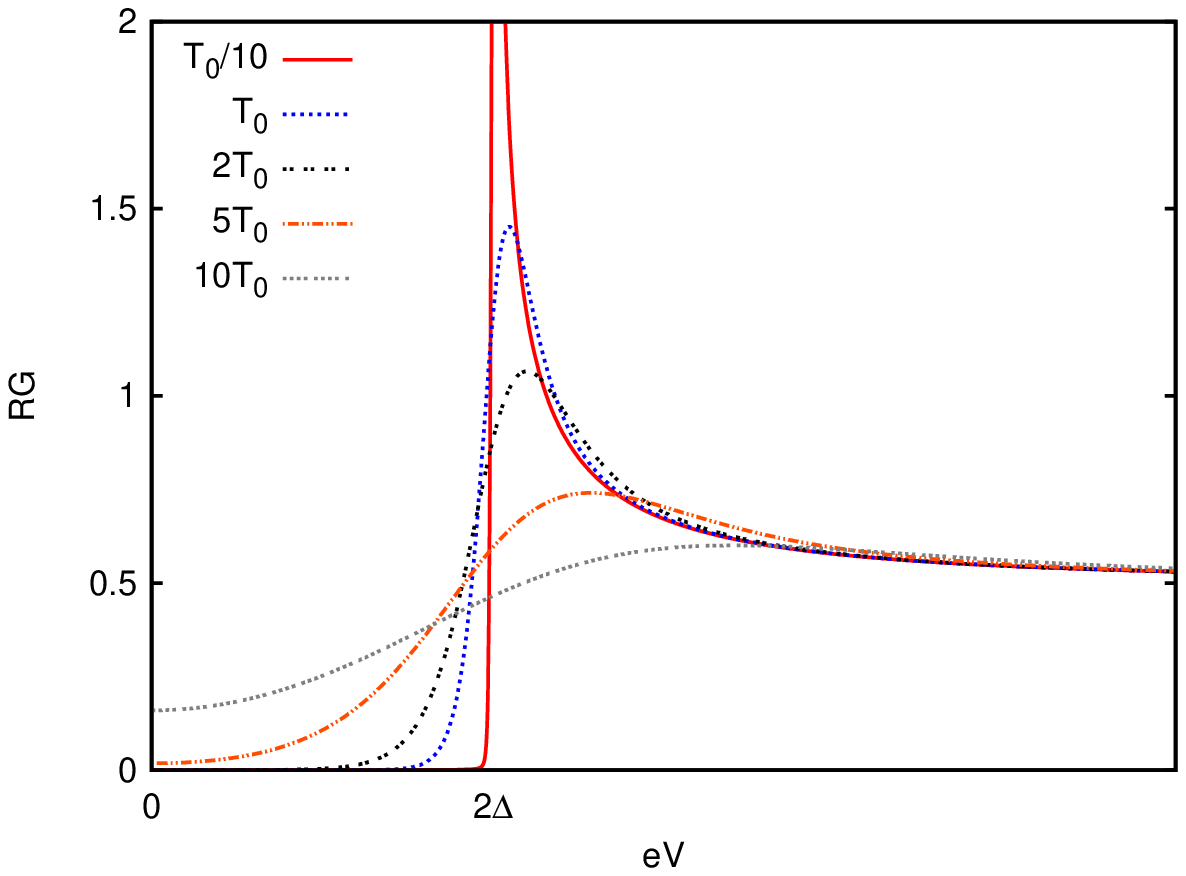}\\
  \includegraphics[width=8cm]{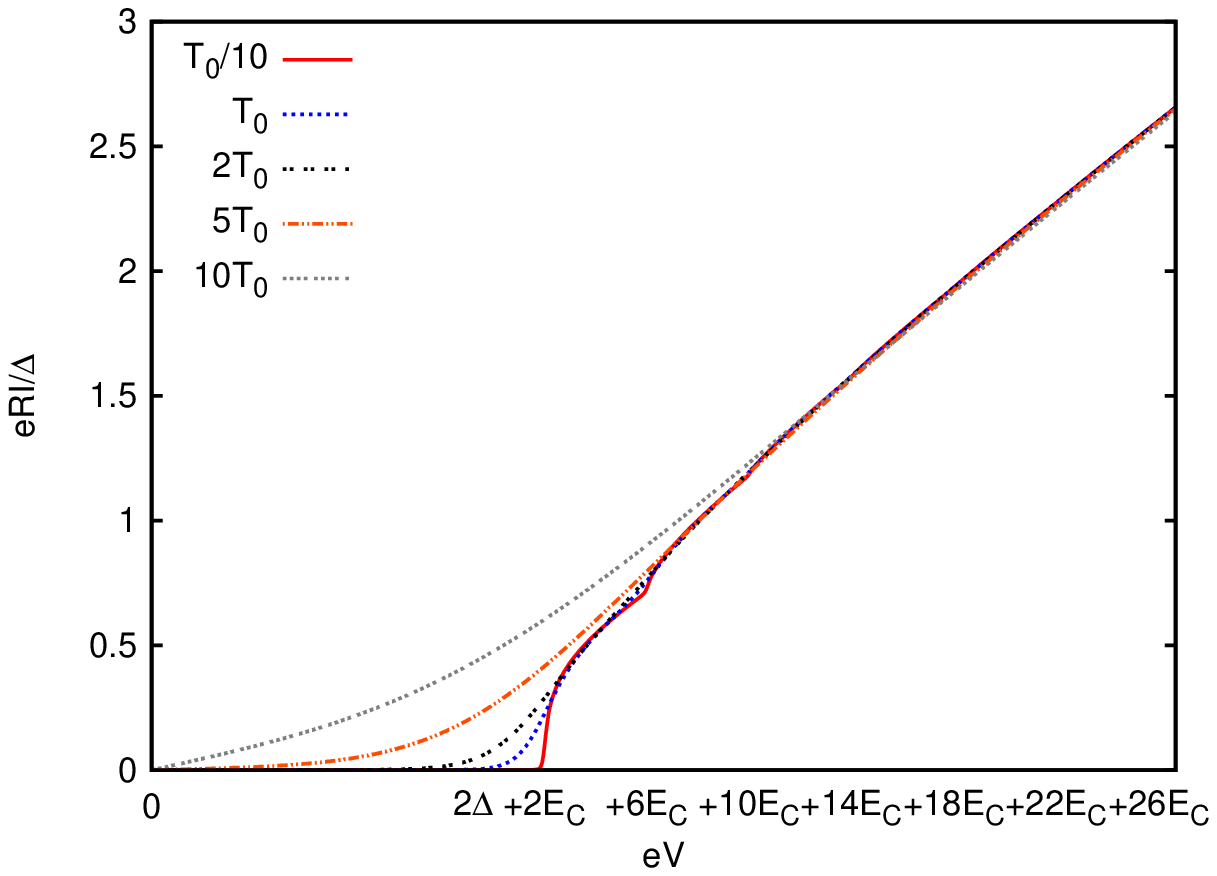}
  \includegraphics[width=8cm]{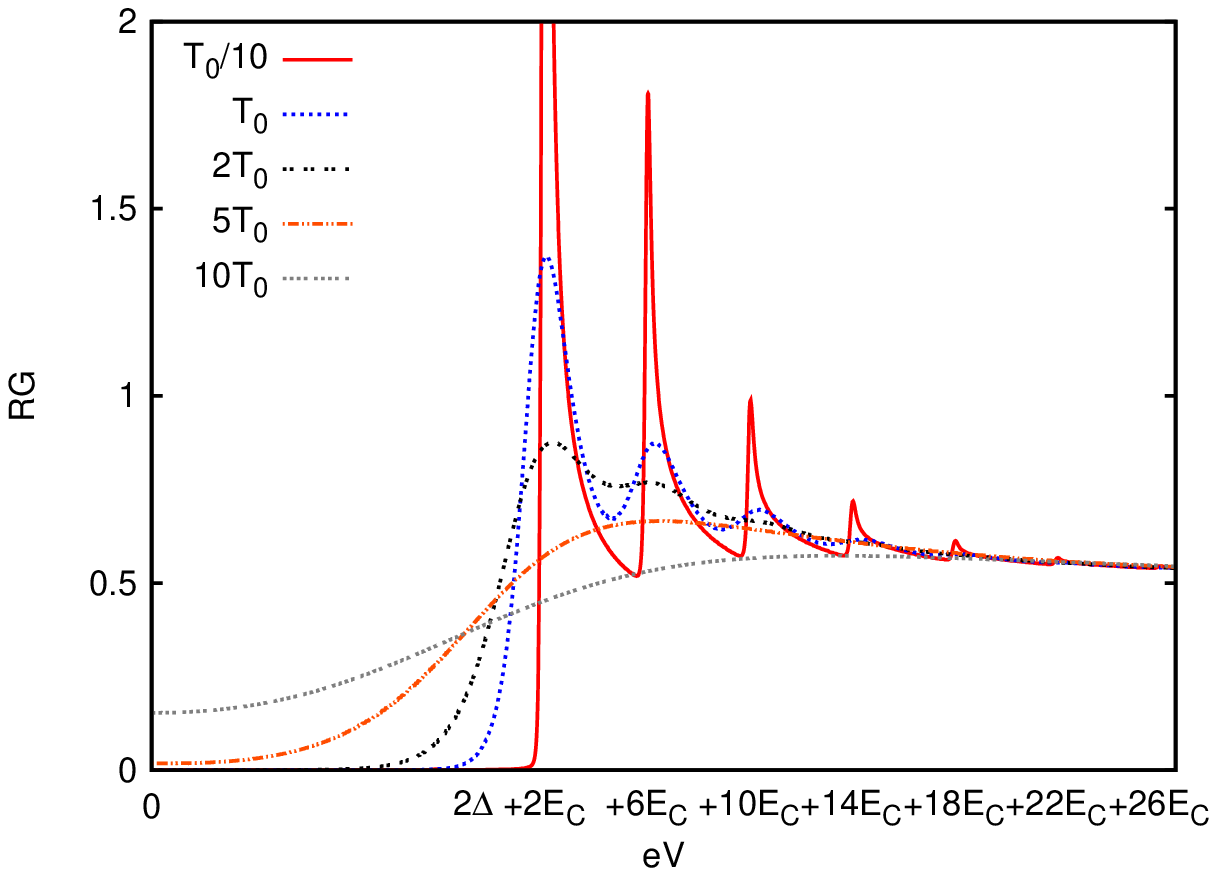}
\end{center}
\caption{Top: Current (left) and conductance (right) of a SINIS structure as function of applied voltage
         at different operating temperatures. The charging energy of the system is set to zero.
         Bottom: Current (left) and conductance (right) through a SINIS structure with finite charging
         energy ($E_C=0.15\Delta$). The offset of the quasiparticle threshold is $2E_C$ and the periodicity
         of its repetition is $4E_C$. In all plots the temperature is $k_BT_0=3.92\times10^{-2}\Delta$.
        }\label{fig_sinis}
\end{figure}

Figure \ref{fig_sinis} shows the $I$--$V$-curves and conductances of a SINIS structure at different operating
temperatures for both zero and finite charging energy. Without charging energy, the SINIS reproduces the
$I$--$V$ curve of a single NIS junction, with all features being at twice the voltage, as we expect due to the fact
that only half of the voltage applied to the SINIS drops over a single junction.
With a finite charging energy, the quasiparticle threshold is pushed up from $V=2\Delta/e$ to
$(2\Delta+2E_C)/e$, and is repeated
at $4E_C$ intervals with decreasing amplitude. This is visible especially well in the conductance plots
in Fig.\ \ref{fig_sinis}. At operating temperatures that are comparable to or larger than the charging energy,
all features are smeared out. In Fig.\ \ref{fig_sinis} (for the case that $\Delta=220\mu $ eV and thus $E_C=33\mu $ eV),
the curve with still barely visible oscillations conductance would correspond to a temperature of
$2k_BT_0\approx17\mu $ eV, while in the curve at the next higher temperature of $5k_BT_0\approx43\mu $ eV the
oscillations are smeared out entirely.

\subsection{Superconducting SETs}
\label{sset}
\begin{figure}[t]
\begin{center}
  \includegraphics[width=8cm]{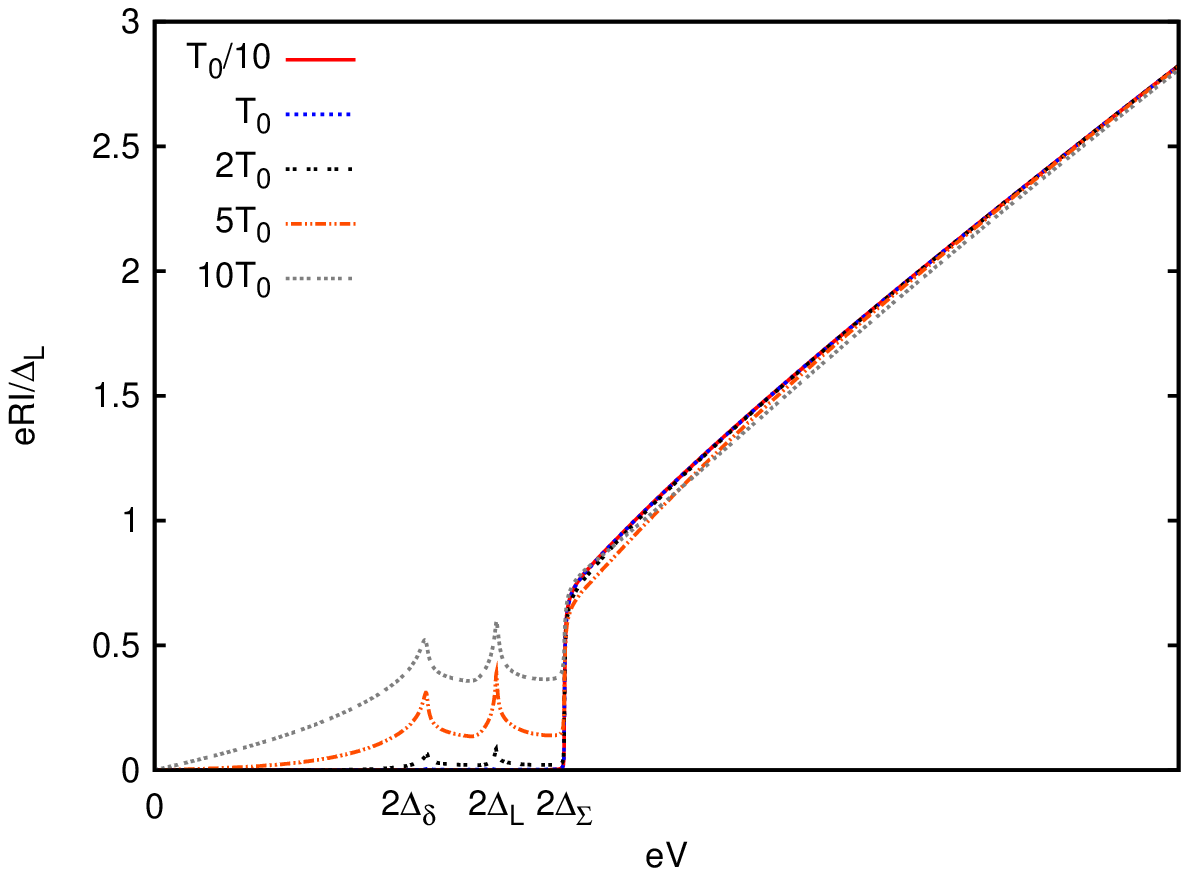}
  \includegraphics[width=8cm]{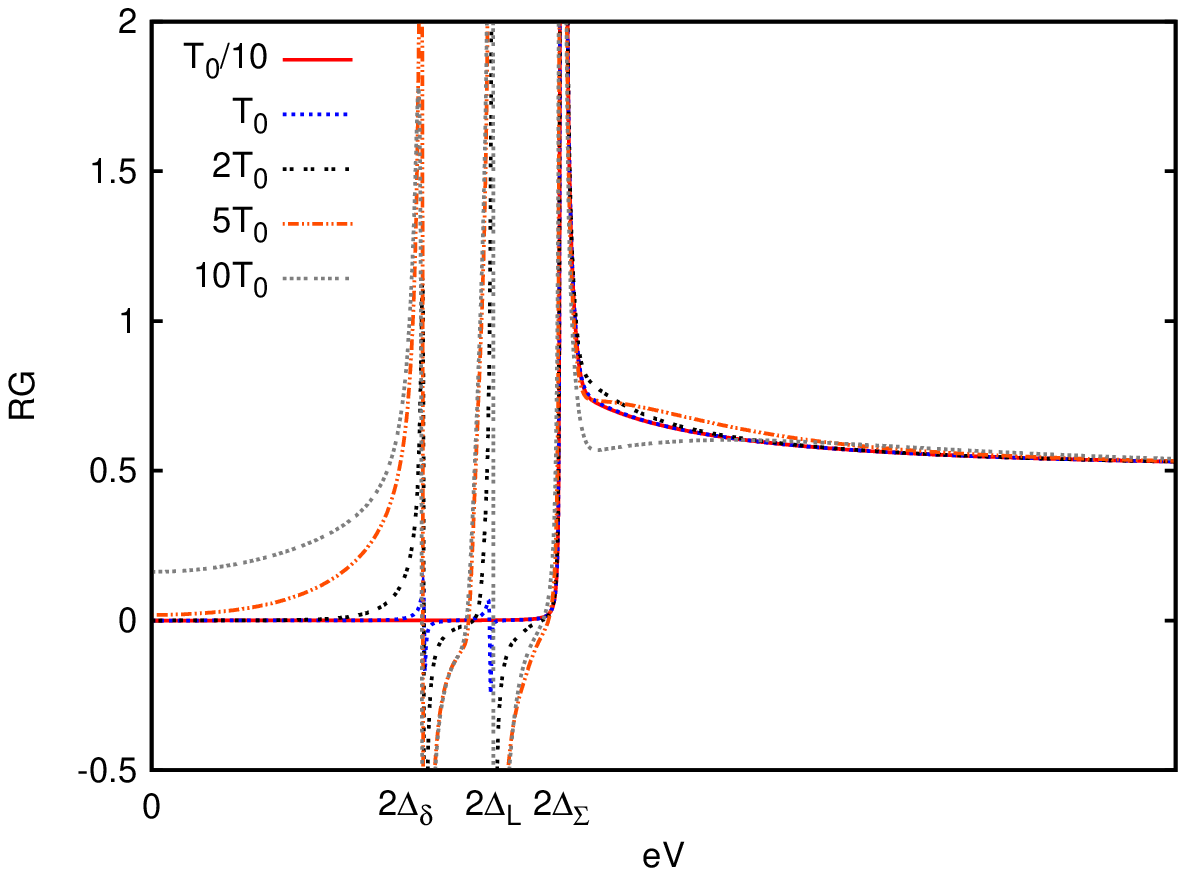}\\
\end{center}
\caption{Current (left) and conductance (right) of an asymmetric SET as function of applied voltage
         at different operating temperatures. The charging energy of the system is set to zero, $k_BT_0=3.92\times10^{-2}\Delta_{\rm L}$,
         and $\Delta_{\rm I}=0.15\Delta_{\rm L}$, where the subscripts $\rm I$ and $\rm L$ denote
         island and lead, respectively.
        }\label{fig_set_sym5_zero_EC}
\end{figure}

When also the island becomes superconducting, we differentiate between two cases, the symmetric SSET, where the
gaps (denoted by $\Delta$) of both leads and island are identical, and the asymmetric SSET,
where the gap in the island differs from the gaps in the leads (denoted by $\Delta_{\rm I}$ and $\Delta_{\rm L}$, respectively).
For simplicity of the further discussion, we introduce two new notations, namely the sum of the lead and island
gaps,
\begin{subequations}
\begin{equation}
\Delta_\Sigma \equiv \Delta_{\rm L}+\Delta_{\rm I}\,,\label{eqn_delta_sigma}
\end{equation}
and the difference between the gaps,
\begin{equation}
\Delta_\delta \equiv |\Delta_{\rm L}-\Delta_{\rm I}|\,.\label{eqn_delta_delta}
\end{equation}
\end{subequations}

Figure \ref{fig_set_sym5_zero_EC} shows the $I$--$V$ characteristics of an asymmetric SET with
$\Delta_{\rm I}=0.2\Delta_{\rm L}$ and
$E_C=0$ for different temperatures.
Compared to the $I$--$V$ of a SINIS structure, we can see the appearance of a new feature,
the singularity-matching peak, which apears at voltages $V^{(sm)}=\pm2\Delta_\delta/e$.
As mentioned above, these peaks appear because of the overlap of two infinite density of states appearing under the
integral giving the tunneling rates.
Singularity-matching peaks in superconducting single-electron transistors have been investigated theoretically previously
in \cite{korotkov} and first observed experimentally by \cite{nakamura} and \cite{pekola}. In  NbAlNb structures
subgap features at $2(\Delta_{\rm Nb} - \Delta_{\rm Al})$ have been observed in \cite{klapwijk}.
Also in this figure we see the usual feature due to quasiparticle threshold at  $V=\pm2\Delta_\Sigma/e$.
Finally, a third peak appears at $V=2\Delta_{\rm L}/e$, where the upper and lower singularity in the density of states
of the left and right lead, respectively, are aligned.
The right plot
in Figure \ref{fig_set_sym5_zero_EC} shows the conductance for the same values. The conductance plot is perhaps
not as descriptive as the one of a SINIS, but it still serves two purposes: 1) it emphasizes very well all sharp
features and thus makes it easy to determine their position and, 2) it allows to easily distinguish
between singularity matching peak- and quasiparticle threshold-like features, as the latter only produces an
upward spike, while the former produces a double spike, with the second spike going downwards. This is especially
useful when discussing SSETs with charging energy, where one can find a large amount of weak features.

At finite charging energies, we see a similar repetition of features
as one sees in SINIS $I$--$V$s with $E_C>0$. These features are a combination of the usual
coulomb blockade features in normal metal SETs and a series of peaks and
steps. The origin of the peaks are the singularity matching peaks in the
tunneling rates (\ref{eqn_tunnel_prob}). Their positions are easily
calculated by equating the energy differences of Eq.\ (\ref{eqn_tunnel_energies})
with the positions of the singularity matching
peaks $\pm\Delta_\delta$, and solving for the applied voltage.

To get a
feeling of  how these features enter the $I$--$V$-curve, we first look at the special
case of two identical junctions, where $C_L=C_R=C_\Sigma/2$.
In this setup, we have that $V_L=-V_R=V/2$. For further simplification, we
also set the gate charge equal to zero, {\it i.e.} $n_g=0$. If we now
for instance look at the series of peaks that correspond to $\delta E_{L\rightarrow I}$, we find
$\pm\Delta_\delta = eV^{\rm (sm)}/2 - 2E_C(n+1/2)$, and therefore
$eV^{\rm (sm)} = \pm 2\Delta_\delta+2E_C(2n+1)$.
Altogether we find that the bias voltages at which peaks occur (at
$n_g=0$) are
\begin{equation}
V^{\rm (sm)}_n =
\pm\frac{2\Delta_\delta}{e}\pm\frac{2E_C}{e}(2n+1)\,.\label{eqn_peaks}
\end{equation}
Similarly, the $I$--$V$ shows a series of positive and negative steps. These
steps originate from the quasiparticle threshold in the tunneling rates
(\ref{eqn_tunnel_prob}). We can find them in a similar manner as the
peaks, by equating Eqs.\ (\ref{eqn_tunnel_energies}) with
$\pm\Delta_\Sigma$ and find
\begin{equation}
V^{\rm (qp)}_n =
\pm\frac{2\Delta_\Sigma}{e}\pm\frac{2E_C}{e}(2n+1)\,.\label{eqn_steps}
\end{equation}
Equation (\ref{eqn_steps}) corresponds to positive steps for
$+\Delta_\Sigma$, and to negative steps
for $-\Delta_\Sigma$. The negative steps however, which are already small
in the tunneling rates, are difficult to
observe in practice.
\begin{figure}[t]
\begin{center}
  \includegraphics[width=8cm]{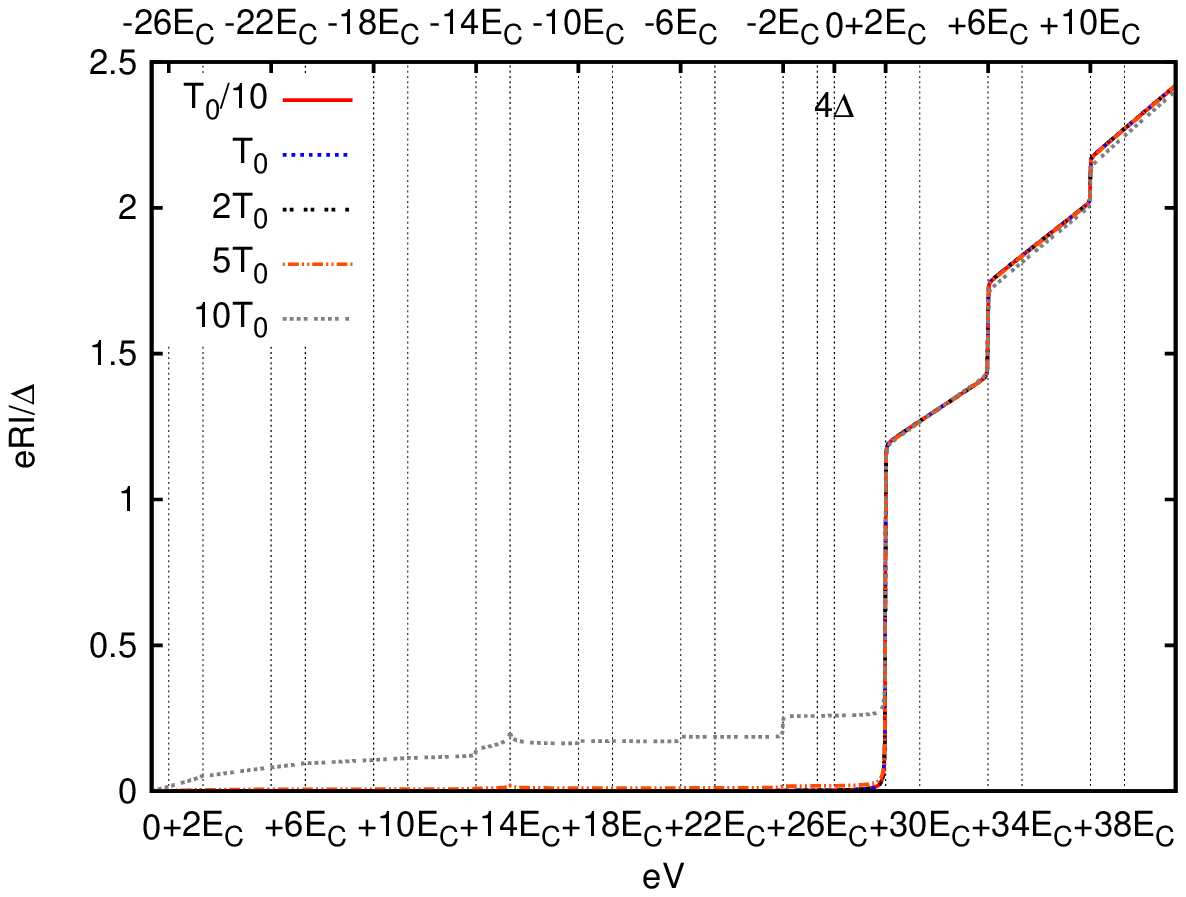}
  \includegraphics[width=8cm]{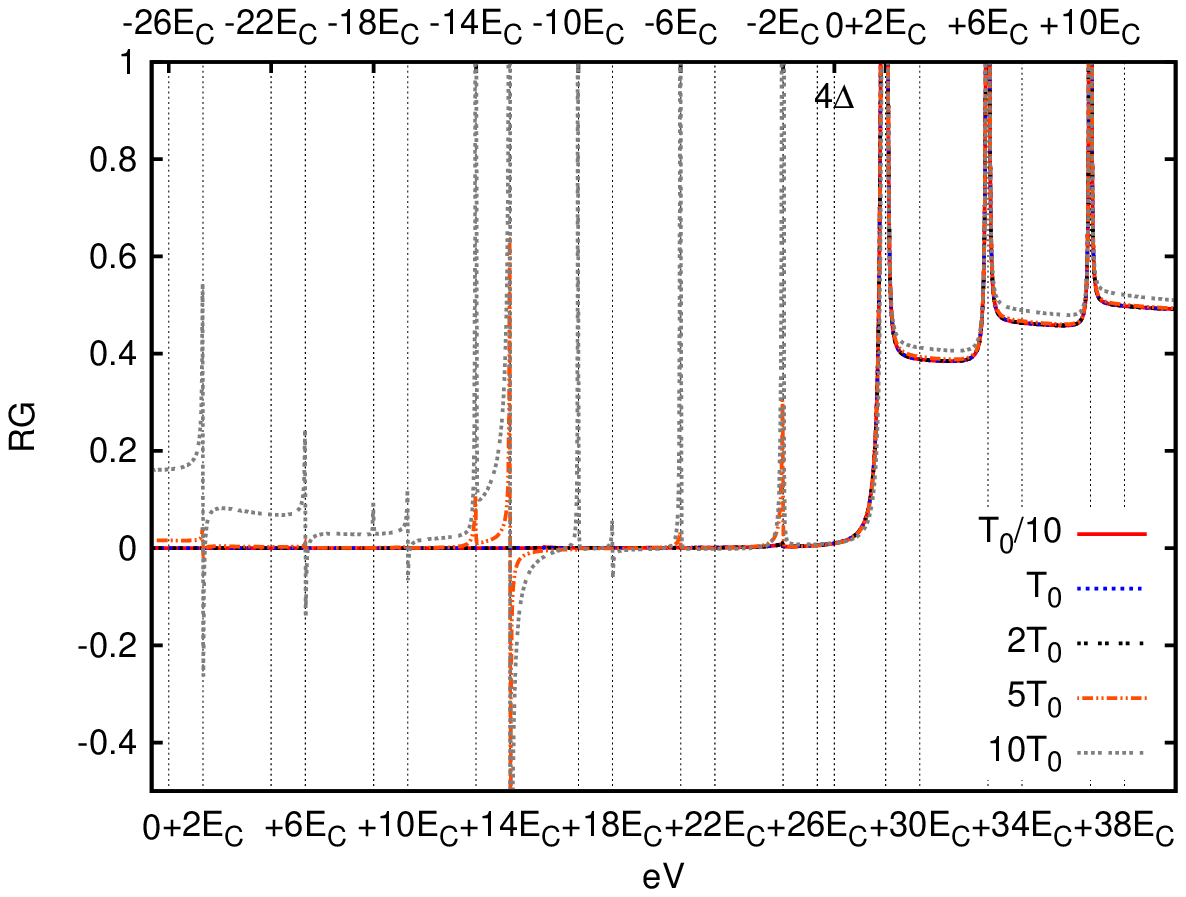}\\
\end{center}
\caption{Current (left) and conductance (right) of a symmetric SET as function of applied voltage
         at different operating temperatures. The gaps are chosen such that
         $\Delta_{\rm I}=\Delta_{\rm L}\equiv\Delta$, and the charging energy of the system is set to $E_C=0.15\Delta$.
         As before, $k_BT_0=3.92\times10^{-2}\Delta$. In
         both plots the top most scale denotes offsets from the value of $eV=4\Delta$.
        }\label{fig_set_sym1}
\end{figure}

\begin{figure}[h]
\begin{center}
  \includegraphics[width=8cm]{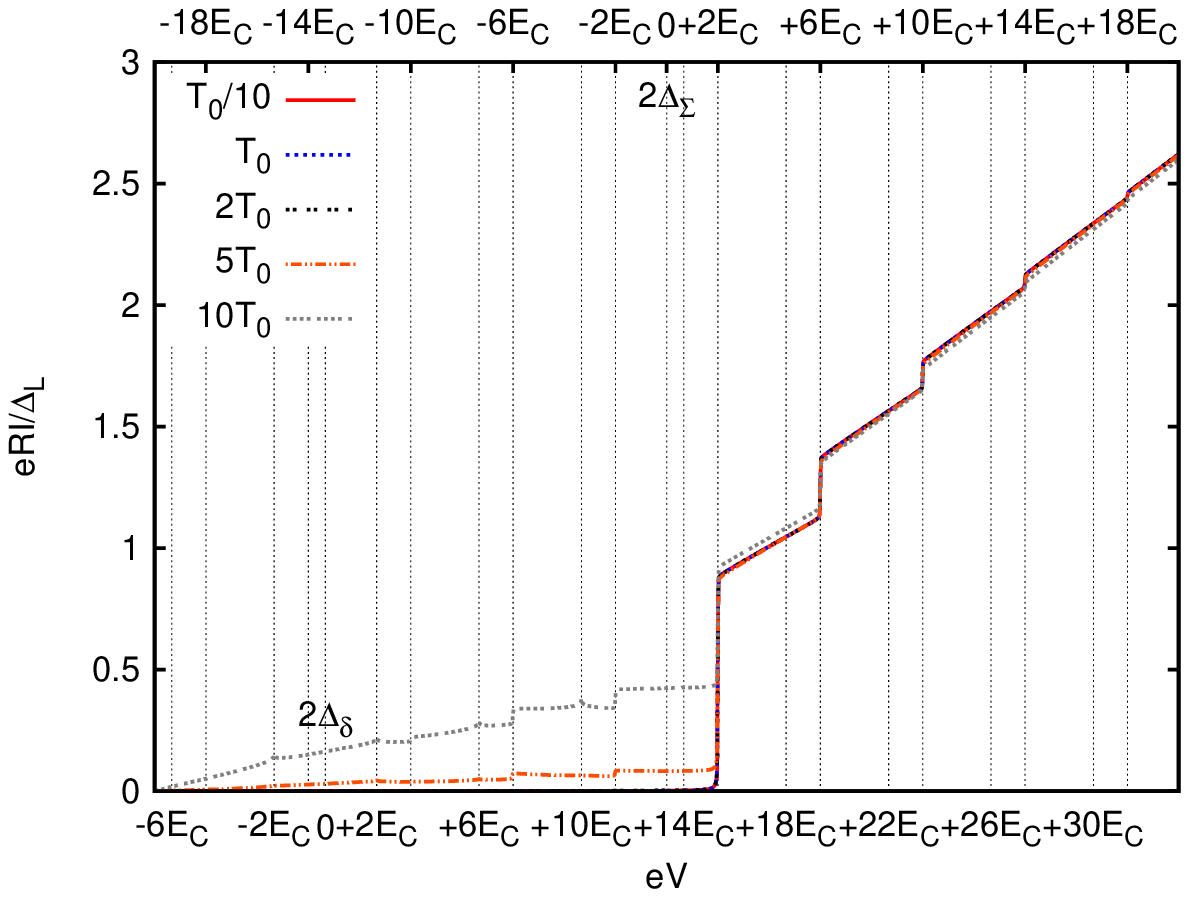}
  \includegraphics[width=8cm]{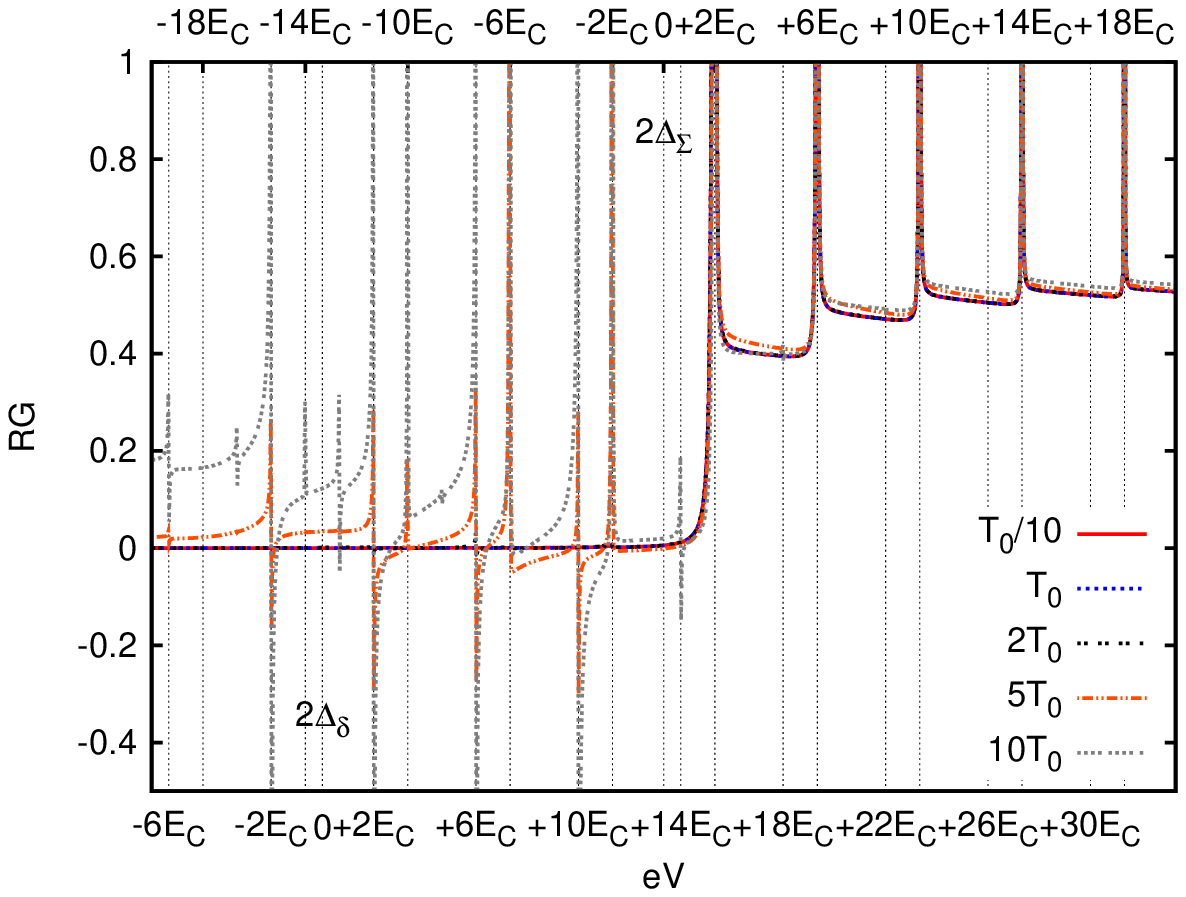}\\
\end{center}
\caption{Current (left) and conductance (right) of an asymmetric SET as function of applied voltage
         at different operating temperatures. The gaps are chosen such that
         $\Delta_{\rm I}=0.5\Delta_{\rm L}$, and the charging energy of the system is set to $E_C=0.15\Delta$.
         As before, $k_BT_0=3.92\times10^{-2}\Delta_{\rm L}$. In
         both plots the top most scale denotes offsets from the value of $eV=2\Delta_\Sigma$,
         while the bottom most scale denotes offsets from the value of $eV=2\Delta_\delta$. The three
         unmarked peaks correspond to $-2\Delta_\delta+10E_C$, $14E_C$ and $18E_C$, respectively.
        }\label{fig_set_sym2}
\end{figure}

\begin{figure}[h]
\begin{center}
  \includegraphics[width=8cm]{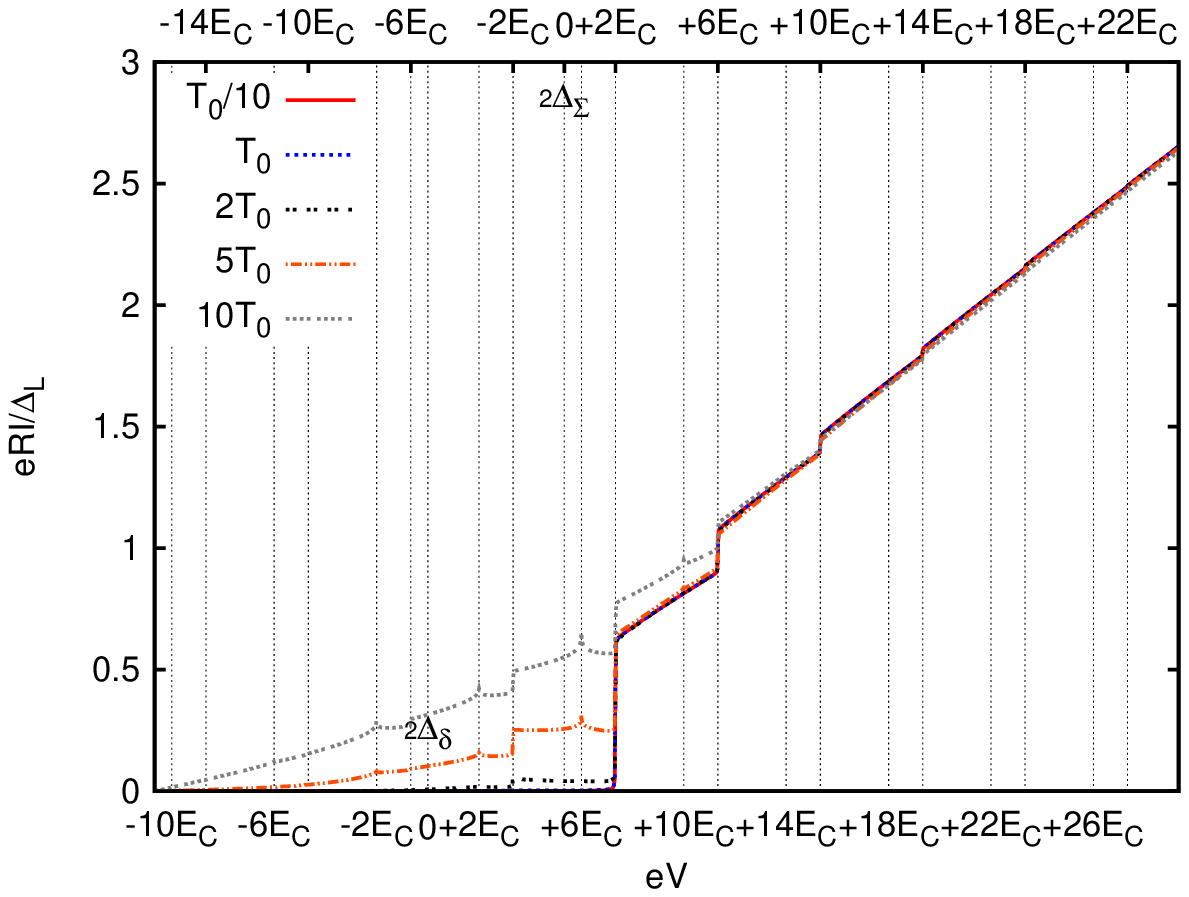}
  \includegraphics[width=8cm]{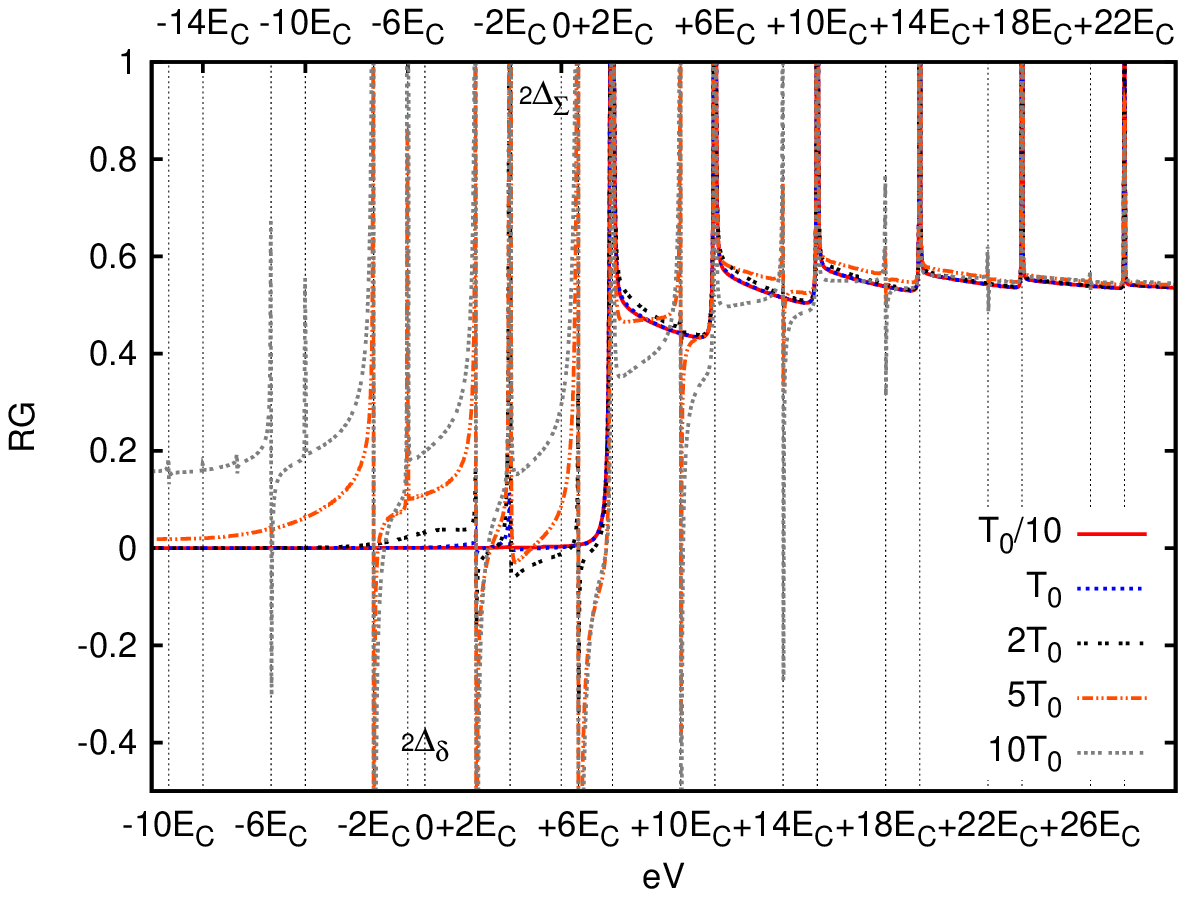}\\
\end{center}
\caption{Current (left) and conductance (right) of an asymmetric SET as function of applied voltage
         at different operating temperatures. The gaps are chosen such that
         $\Delta_{\rm I}=0.2\Delta_{\rm L}$, the charging energy of the system is set to $E_C=0.15\Delta$,
         and $k_BT_0=3.92\times10^{-2}\Delta_{\rm L}$. In
         both plots the top most scale denotes offsets from the value of $eV=2\Delta_\Sigma$,
         while the bottom most scale denotes offsets from the value of $eV=2\delta_\Delta$.The
         unmarke peaks correspond to $-2\Delta_\delta+14E_C$.
        }\label{fig_set_sym5}
\end{figure}

Figures \ref{fig_set_sym1}, \ref{fig_set_sym2} and \ref{fig_set_sym5} show the $I$--$V$ and conductance
of SSETs with a finite charging energy of $E_C=0.15\Delta_L$, with ratios between island and lead gaps
of  $\Delta_{\rm I}/\Delta_{\rm L}=$ $1$, $0.5$ and $0.2$, respectively. In all figures one can see the distribution
of peaks and steps as described in Eqs.\ (\ref{eqn_peaks}) and (\ref{eqn_steps}), with growing complexity.
In the $I$--$V$ curves, especially in Figure \ \ref{fig_set_sym1}, the features are difficult to distinguish. Therefore
the conductance plots become a powerful tool in identifying the position of peaks and steps, and in distinguishing
between them (single and double spikes in conductance correspond to steps and peaks, respectively, in current).
\subsection{Effect of the gate}
\label{gate}
\begin{figure}[t]
\begin{center}
  \includegraphics[width=8cm]{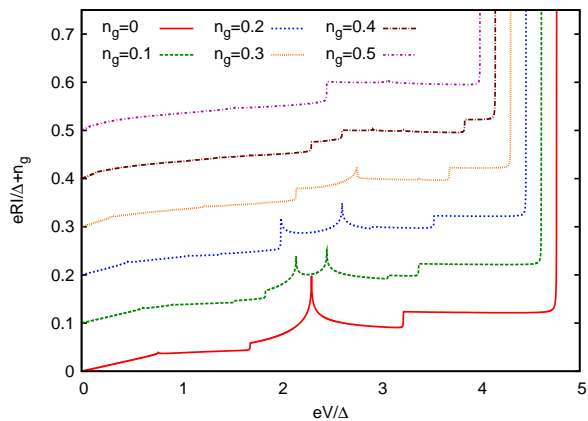}
\end{center}
\caption{The effect of the gate charge. Here $\Delta=2.6E_C$ and
         $T=0.8E_C$. The curves are offset by the respective gate charge for better distinguishability.}
         \label{fig_nakamura}
\end{figure}

To study how the gate charge affects the IV-curves of an SSET, it is best
to do some simplifications again, to somewhat limit the amount of
confusing details. In Fig.\ \ref{fig_nakamura} we have reproduced the
first figure of \cite{nakamura}, where the IV of
a symmetric SSET was studied. The figure shows a set of IV curves through
the SSET as function of bias voltage for different values of the gate
charge. As one can see, also here appears a series of peaks and steps, but
their position is shifted as the gate charge is altered. Again the
position of the features is easily calculated by equating the energy
differences (for the most general case, use
(\ref{eqn_tunnel_energies_direct})) to $\pm\Delta_\delta$ for peaks, and
$\pm\Delta_\Sigma$ for steps. As in this case $\Delta_\delta=0$, the
equation for the position of the peaks is somewhat simplified, as here the
derived voltages will not depend on the superconducting gap.
As in \cite{nakamura}, the positions of the
peaks in the system at hand (where possibly $C_{\rm L}\ne C_{\rm R}$)
are given by
\begin{equation}
V_{\rm \{L,R\},n}^{\rm sm} =
\frac{eC_{\rm \{L,R\}}}{C_{\rm L}C_{\rm R}}\left[\frac{1}{2} \pm \left(n+n_g^\prime\right)\right]\,. \label{sto}
\end{equation}
Here we defined $n_g^\prime\equiv n_g+Q_0/e$, with
$en_g$ the gate charge and $Q_0$ the background charge of the island.
Accordingly, the positions of the positive and negative steps are given by
\begin{equation}
V_{\{\rm L,R\},n}^{\rm qp } =
\frac{eC_{\rm \{L,R\}}}{C_{\rm L}C_{\rm R}}\left[\pm \frac{2\Delta C_\Sigma}{e^2}+
\frac{1}{2}\pm \left(n+n_g^\prime\right)\right]\,. \label{stur}
\end{equation}
%
%
%
Here the positive steps are  obtained for the  $+$ sign in from of the fraction $2\Delta C_\Sigma /e^2$, and the negative steps are
obtained for a $-$ sign in front of the same fraction. The negative steps can be observed only at higher temperatures -- see the comments on the height of the quasiparticle
threshold, Eq. (\ref{eqn_finite_jumps_a}) and (\ref{eqn_finite_jumps_b}); only the positive steps are visible
in Figure (\ref{fig_nakamura}).

It is easy to verify that Eqs.\ (\ref{eqn_peaks}) and (\ref{eqn_steps})
are another case of the expressions derived above Eqs. (\ref{stur}-\ref{sto}); these sets of equations become identical
when using the same
assumptions (identical junctions and $n_g^\prime$=0), and remembering that here
$\Delta_\delta=0$ and $\Delta_{I}=\Delta_{L}=\Delta_{R}=\Delta$.

\subsection{Cooling}
\label{cooling}
Tunneling of electrons through junctions results in energy being transferred between the electrodes, as the higher energetic electrons are extracted or pumped from or into one of the electrodes. In the case of two-junction structures such
as SETs, SINIS, {\it etc.}, this results in the heating and cooling of the middle electrode (the island). This is due to the fact that, although
the currents through the two junctions are the same, as it should be due to the conservation of charge (the charge of the island has to remain constant)
the energy transfer (heat) does not need to satisfy such a conservation law.

It is straightforward to calculate the energy transferred per unit time out of the island (cooling power) as a function of the applied voltage,
\begin{equation}
P_{\rm I}(V) = \sum\limits_{n=-\infty}^{\infty}p(n)
\left[P_{\rm I\rightarrow L}(n)-P_{\rm L\rightarrow I}(n)+P_{\rm I\rightarrow R}(n)
-P_{\rm R\rightarrow I}(n)\right]\,.\label{eqn_diss_def}
\end{equation}
Here the powers $P_{\rm L\rightarrow I}(n)$, $P_{\rm I\rightarrow L}(n)$,
$P_{\rm R\rightarrow I}(n)$, and $P_{\rm I\rightarrow R}(n)$ correspond to energy being transferred
onto/off the island by tunneling into the left and right electrodes, at a given number of excess electrons
$n$ on the island.
They can be calculated by multiplying the energy carried by each tunneling electron to the probability of
tunneling per unit time (given by the density of states
and Fermi factors), and then summing over
energy states (see also Eqs. (\ref{eqn_tunnel_prob_a}), (\ref{eqn_tunnel_prob_b})),
\begin{subequations}
\begin{eqnarray}
P_{\rm I\rightarrow X}(n)
 &=& \frac{1}{e^2R_{IX}}\lint{-\infty}{\infty}{E}EN_{\rm I}(E)N_{\rm X}[E+\delta E_{\rm I\rightarrow X}(n)]f_{\rm I}(E)\left[1-f_{\rm X}(E+\delta E_{\rm I\rightarrow X}(n))\right]
     \label{eqn_diss_integral_a}\\
P_{\rm X\rightarrow I}(n)
 &=& \frac{1}{e^2R_{XI}}\lint{-\infty}{\infty}{E}\left[E+\delta E_{\rm X\rightarrow I}(n)\right]
     N_{\rm I}\left[E+\delta E_{\rm X\rightarrow I}(n)\right]
     N_{\rm X}(E)\left[1-f_{\rm I}(E+\delta E_{\rm X\rightarrow I}(n))\right]f_{\rm X}(E)
     \label{eqn_diss_integral_b}\,,
\end{eqnarray}
\end{subequations}\label{eqn_diss_integral}
where $\rm X=L,R$.
\begin{figure}[t]
\begin{center}
  \includegraphics[width=8cm]{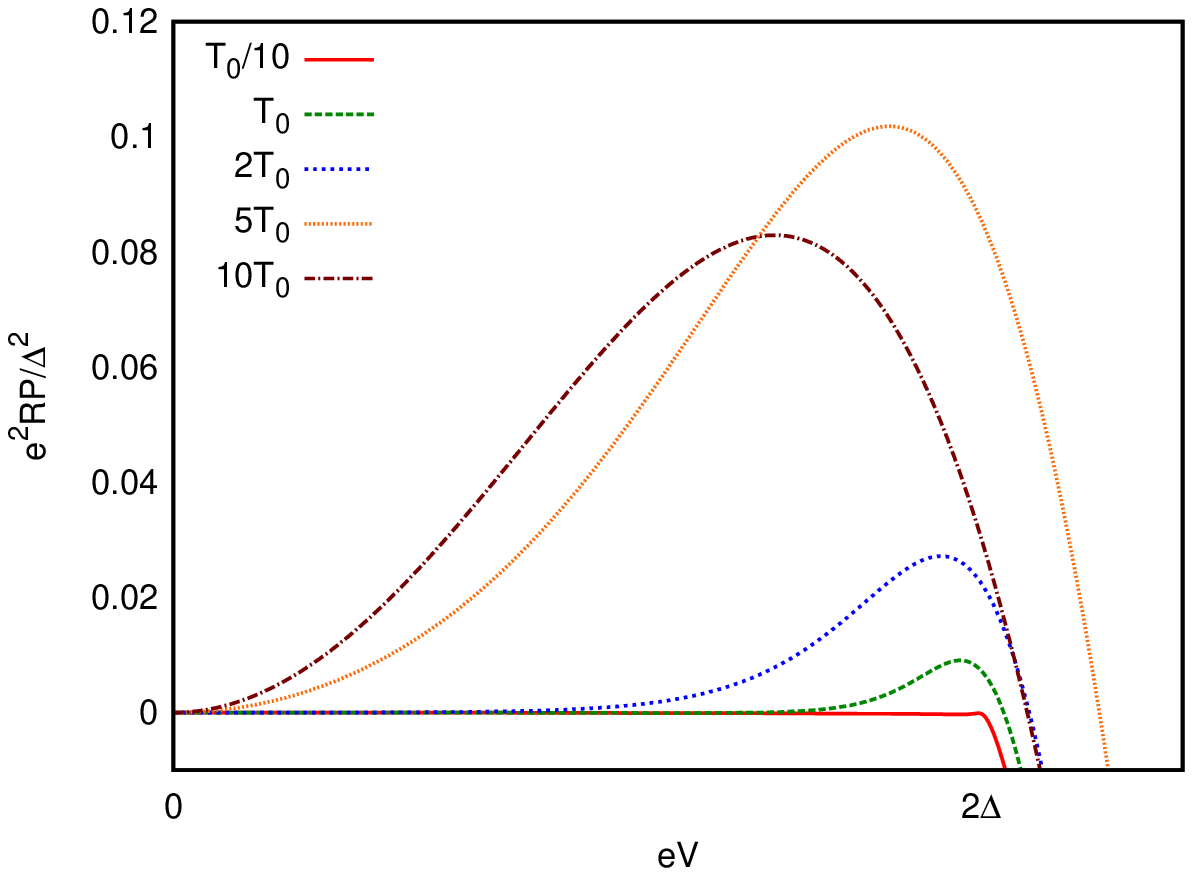}
  \includegraphics[width=8cm]{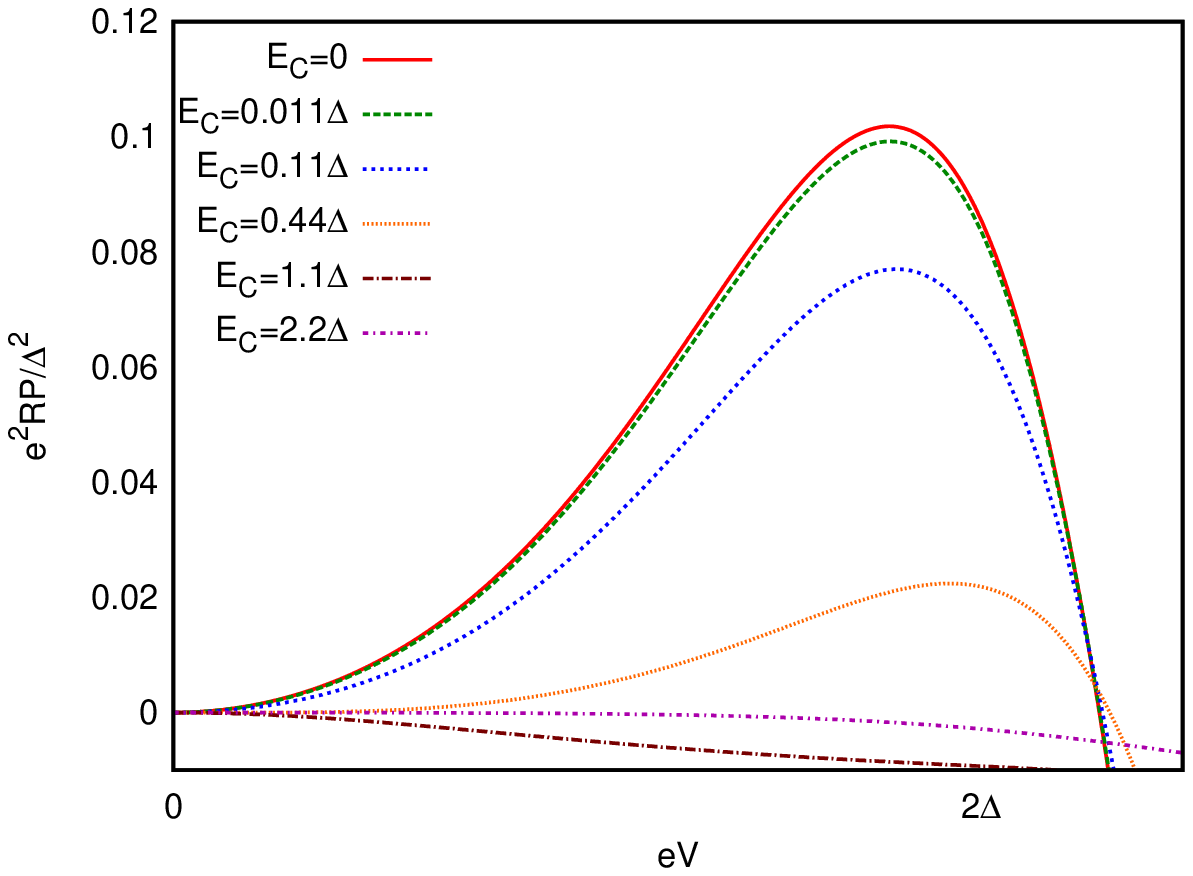}\\
  \includegraphics[width=8cm]{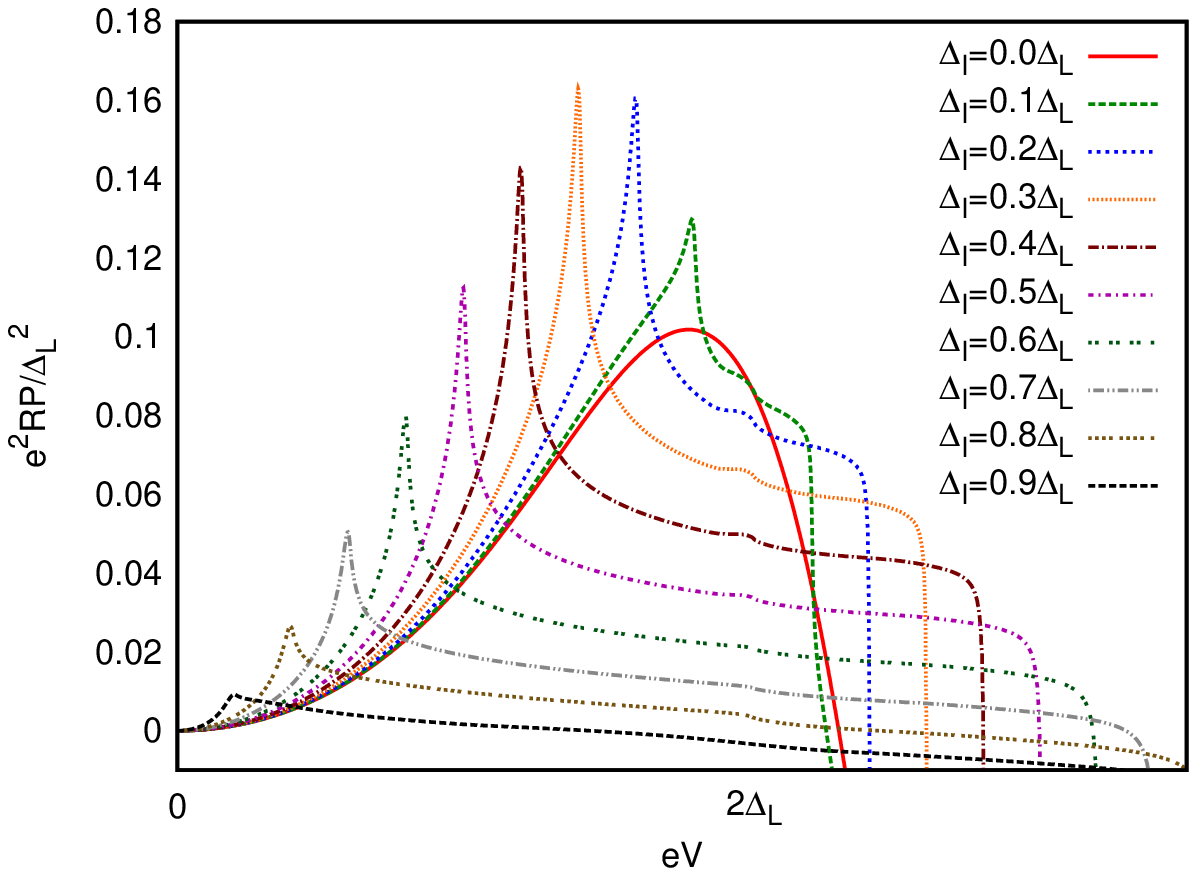}
  \includegraphics[width=8cm]{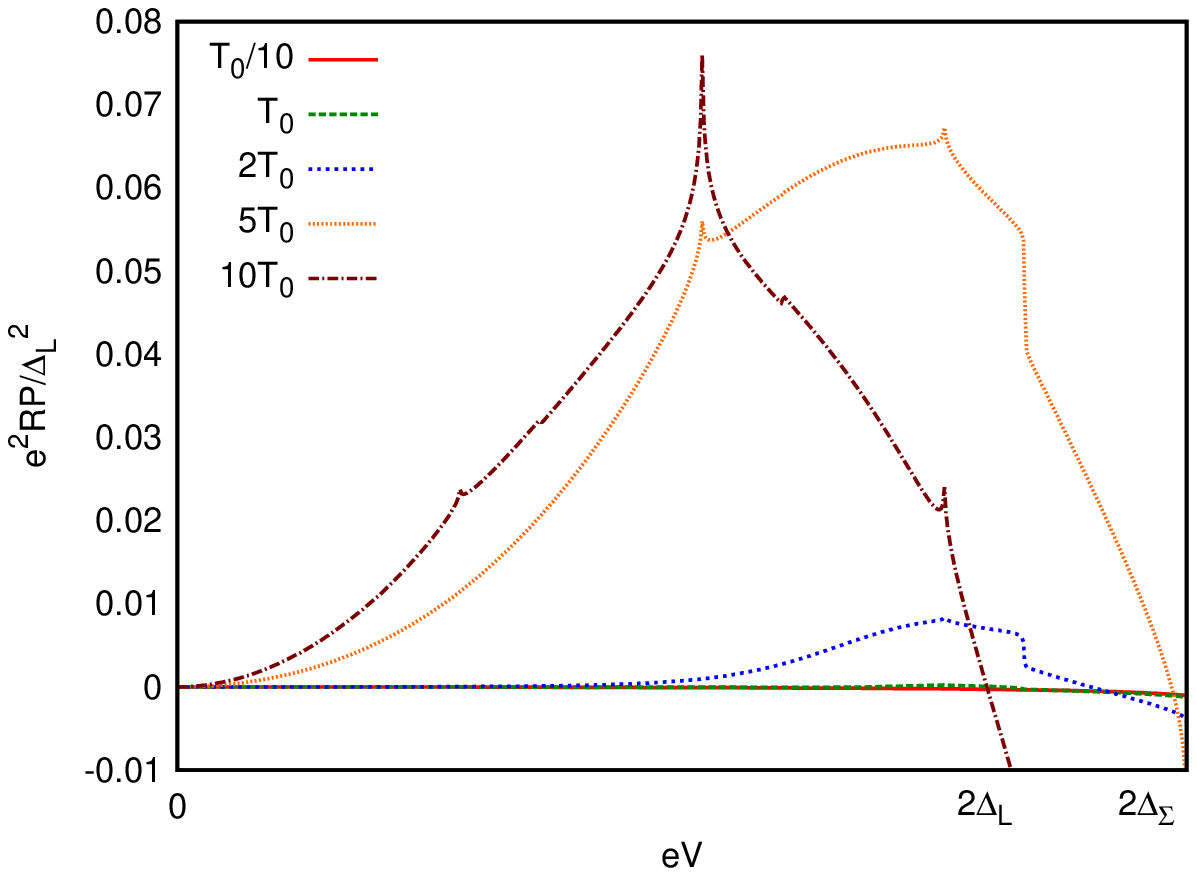}
\end{center}
\caption{Cooling powers in different double-island devices. Top left:  SINIS structure with $E_C=0$ at different
temperatures. Top right: SINIS at fixed temperature $5T_0$ with different charging energies. Bottom left: SET
with $5T_0$ and $E_C=0$ for different gap sizes $\Delta_{\rm I} < \Delta_{\rm L}$ of the island. Bottom right:
SET with $\Delta_{\rm I}=0.2\Delta_{\rm L}$ and $E_C=0.15\Delta_{\rm L}$ at different temperatures.
In all plots $k_BT_0=3.92\times10^{-2}\Delta_{\rm L}$.
}\label{fig_cooling}
\end{figure}

In Figure \ref{fig_cooling} we present a comparison between the cooling powers on the island Eq.(\ref{eqn_diss_def})
for SINIS (upper plots) and superconducting SET structures (lower plots). We observe that cooling exists only in an
interval of bias voltages not much larger than the quasiparticle threshold at $2\Delta /e$ (we take identical
superconducting leads with gap $\Delta$). As seen from the left upper plot, as the temperature is lowered the
process becomes much less efficient. The general effect of the charging energy is that it is detrimental to cooling
(right plot, upper part). In the case of superconducting SET's (lower plots, Figure \ref{fig_cooling})
the existence of singularity-matching peaks produces relatively sharp spikes in the cooling power. Essentially,
the island is cooled down by BCS quasiparticles \cite{manninen}. The left figure shows these features for the case of
negligible charging energy (a SISIS structure with a relatively large middle electrode having a different gap than
the leads). An interesting feature is also that the
range of voltages over which cooling occurs is extended, due to the fact that now the quasiparticle threshold is
at values $2\Delta_{\rm L}+2\Delta_{\rm I}$. Finally, cooling in SET structures with finite charging energy is
shown in the lower-right plot of Figure \ref{fig_cooling}.

\section{Conclusions}
\label{conclu}

We have presented the theory of tunneling in metallic and superconducting two-junction arrays such as single electron
transistors, together with a number of applications.
All three energy scales, the charging energy, the superconducting gap, and the temperature, are considered and their
role is thoroughly discussed. For example, we examined how a finite superconducting gap affects the Coulomb-blockade
based thermometry, the effect of singularity-matching peaks in the current-voltage and conductance-voltage
characteristics of
superconducting single-electron transistors, and we looked at the effect of charging energy in cooling devices. With the development of the field of nanotechnology, such devices could emerge as very useful tools for
high-precision measurements of nano-structured materials and objects at low temperatures.

\appendix

\section{A useful integral}\label{sec_useful}

For completeness, we give a full derivation of the result Eq. (\ref{rrrt}) (see {\it e.g.} \cite{ingold}). The Fermi functions are given by
\begin{equation}
f(\epsilon ) = \frac{1}{\exp (\epsilon /k_{B}T)+1}.
\end{equation}
We now introduce the function
\begin{equation}
g(x) = \int\limits_{-\infty}^{\infty} d\epsilon\left[f(\epsilon )-f(\epsilon +x)\right]
\end{equation}
Obviously $g(0)=0$. The first derivative of $g$ gives
\begin{equation}
\frac{dg(x)}{dx}=-\int\limits_{-\infty}^{\infty}d(\epsilon +x) \frac{df(\epsilon + x)}{d(\epsilon +x)} =
-f(\infty )+f(-\infty ) = 1.
\end{equation}
This means that $g(x) = x$, so we have the result
\begin{equation}
\int\limits_{-\infty}^{\infty} d\epsilon\left[f(\epsilon )-f(\epsilon +x)\right] =x\,.\label{ap}
\end{equation}
Now, integrals over expressions of the type $f(E)[1-f(E+x)]$, which appear
in the theory of tunneling due to Pauli exclusion principle, can be solved
by noticing that
\begin{equation}
f(\epsilon )[1-f(\epsilon +x)]=\frac{f(\epsilon )-f(\epsilon +x)}{1-\exp (-\beta x)},
\end{equation}
where $\beta = 1/k_{B}T$.
Now, using Eq. (\ref{ap}) we get immediately
\begin{equation}
\int_{-\infty}^{\infty} d\epsilon f(\epsilon)[1-f(\epsilon +x)] = \frac{x}{1-\exp (-\beta x)}.
\end{equation}

\section{Derivation of some integral expansions used}\label{sec_expansions}
{\it Generic procedure:} In the following we want to derive the expansion of Eq.\ (\ref{eqn_expansion_of_Upsilon}).
 To get a feeling for the procedure, we will
first consider a somewhat easier and more common integral, namely
\begin{eqnarray}
\Upsilon (c)
 &=& \lint{0}{\infty}{x}\frac{1}{x+c}h(x)\,,\label{eqn_divergent_integral}
\end{eqnarray}
which we assume to not be solvable analytically.
For the function $h(x)$ we assume that it is well behaved for all $x\in [0,\infty)$ and
that $h(x)$ behaves as $\propto\exp(-x)$ for large $x$. However, $\Upsilon(c)$ is logarithmically divergent for
$c=0$. We can separate $\Upsilon(c)$ into a divergent and an non-divergent part by performing a partial integration,
\begin{eqnarray}
\Upsilon (c)
 &=& \ln(1/c)h(0)-\lint{0}{\infty}{x}\ln(x+c)h^\prime(x)\label{eqn_ln_integral}\,.
\end{eqnarray}
Here the first term is an explicit expression that describes the divergence around $c=0$ and the second term is a finite
integral that needs to be computed numerically (we use the notation
$h^\prime(x)\equiv dh(x)/dx$). Unfortunately, the function over the integral in the second term of
Eq.\ (\ref{eqn_ln_integral}) has no Taylor expansion in $x+c\rightarrow 0$, making any further analysis of the integral
difficult. It is however possible to find
an expansion by doing repeated partial integration.
If we define the function $F_n(x)$ such that
\begin{eqnarray}
\frac{d^n}{dx^x}F_n(x)
 &=& \ln(x)\,,
\end{eqnarray}
it is easy to show that
\begin{eqnarray}
F_n(x)
 &=& \frac{1}{n!}x^n[\ln(x)-H_n ]\,, \label{eqn_F_n}
\end{eqnarray}
where $H_n=\sum_{k=1}^n1/k$ is the harmonic number (we could actually still add an arbitrary polynomial of $n$th order
to $F_n$, but due to the repeated partial integrations all its coefficients cancel out and we therefore set them
equal zero). These functions have the properties $F_0(x)=\ln(x)$,
$F_n(x=0)=0, \,\,\left(\forall\right)\,n>0$ and
$F_{n\rightarrow\infty}(x)\rightarrow0, \,\,\left(\forall\right)\,x>0$. With this we can write $\Upsilon (c)$ as
\begin{eqnarray}
\Upsilon (c)
 &=& \ln(1/c)h(0)-\sum_{n=1}^\infty(-1)^n\frac{1}{n!}c^n[\ln(c)-H_n]h^{(n)}(0)\,,
\end{eqnarray}
where by $h^{(n)}(x)$ we denote the $n$th derivative of $h(x)$, $d^nh(x)/dx^n$.

\vspace{3mm}

{\it Application:} In Section \ref{sec_SIS}, we have encountered slightly more complicated integrals, which were of the form
\begin{eqnarray}
\Upsilon (c)
 &=& \lint{0}{\infty}{x}\frac{1}{\sqrt{x (x+c)}}h(x)\,,
\end{eqnarray}
with a logarithmic singularity in $c=0$. For this reason we would like to expand $\Upsilon^\prime (c)$
after the recipe that
was used when expanding the previous integral, Eq. (\ref{eqn_divergent_integral}). The only complication we encounter is that we need to find a function $G_n(x)$ with
$dG_0(x)/dx=1/\sqrt{x(x+a)}$ and
\begin{eqnarray}
\frac{d^n}{dx^n}G_n(x)
 &=& G_0(x)\,.\label{eqn_G_n_consistency}
\end{eqnarray}
It is easy to verify that the function $G_0$ is given by
\begin{eqnarray}
G_0(x)
 &=& 2\ln(\sqrt{x}+\sqrt{x+c}) .
\end{eqnarray}
The general function $G_n$ can be found with the trial solution
\begin{eqnarray}
G_n(x)
 &=& P_n(x)\ln(\sqrt{x}+\sqrt{x+c})+Q_n(x)\sqrt{x(x+c)}\,,\label{eqn_G_n_ansatz}
\end{eqnarray}
where $P_n(x)$ and $Q_n(x)$ are polynomials. By inserting Eq.\ (\ref{eqn_G_n_ansatz}) into
Eq.\ (\ref{eqn_G_n_consistency}), we find that $P_n$ and $Q_n$ must fulfill the relations
\begin{eqnarray}
\frac{dP_n(x)}{dx}
 &=& P_{n-1}(x)\,,
\end{eqnarray}
and
\begin{eqnarray}
P_n(x)+(2x+c)Q_n(x)+2x(x+c)\frac{dQ_n(x)}{dx}
 &=& 2x(x+c)Q_{n-1}(x)\,.
\end{eqnarray}
These relations can finally shown to be fulfilled if
\begin{subequations}
\begin{eqnarray}
P_n(x)
 &=& \sum_{k=0}^{n} \frac{1}{k!}\alpha_{n-k}x^k\\
Q_n(x)
 &=& \sum_{k=0}^{n-1}\beta_k^nx^k\,,
\end{eqnarray}
\end{subequations}
where
\begin{subequations}
\begin{eqnarray}
\alpha_n
 &=& \frac{2}{4^n}\frac{(2n)!}{(n!)^3}c^n\\
\beta_k^n
 &=& \sum_{l=0}^{k}\left(\frac{-1}{c}\right)^{l+1}\frac{4^l(l!)^2}{(2l+1)!(k-l)!}\alpha_{n-k+l}\,.
\end{eqnarray}
\end{subequations}
Although these polynomials look somewhat complicated, only the constant term in $P_n(x)$ will enter the expansion
of $\Upsilon (c)$,
\begin{eqnarray}
\Upsilon (c)
 &=& -\sum_{n=0}^{\infty}G_n(0)h^{(n)}(0)\nonumber\\
 &=& -\sum_{n=0}^{\infty}\alpha_n\ln(\sqrt{c})h^{(n)}(0)\nonumber\\
 &=& \ln\left(\frac{1}{c}\right)h(0)-\sum_{n=1}^{\infty}\frac{(2n)!}{4^n(n!)^3}c^n\ln(c)h^{(n)}(0),\label{eqn_final_expansion}
\end{eqnarray}
where the first term of Eq.\ (\ref{eqn_final_expansion}) contains the convergence for $c\rightarrow0$.

The convergence of expression (\ref{eqn_final_expansion}) is somewhat tricky to show as it depends on the
values of the function $h(x)$ and its derivatives at $x=0$. With the ratio test of convergence for infinite
series we find, writing Eq.\ (\ref{eqn_final_expansion}) as $\sum_n a_n$, that
\begin{eqnarray}
\left|\frac{a_{n+1}}{a_n}\right|
 &=& \frac{(2n+1)(2n+2)}{4(n+1)^3}c\frac{h^{(n+1)}(0)}{h^{(n)}(0)}\nonumber\\
 &\rightarrow&\frac{c}{n}\left[\frac{d}{dx}\ln(h^{(n)}(x))\right]_{x=0}\hspace{5mm}\mathrm{for}\hspace{5mm}n\rightarrow\infty\,,
\end{eqnarray}
which is smaller than one if $(d/dx)\ln(h^{(n)}(x))|_{x=0}$ increases slower than $n$. For example, if $h(x)=\exp(-ax)$,
we would have $|a_{n+1}/a_n|\rightarrow ac/n$, which tends to zero for large $n$, and thus the series would be convergent.

\acknowledgments

We would like to thank J. P. Pekola and I. Maasilta for useful comments on the manuscript.
T. K. would like to acknowledge financial support from the Emil Aaltonen foundation.
The contribution of G.S.P. was supported by the Academy of Finland (Acad. Res. Fellowship 00857, and projects 129896, 118122, and 135135).

\end{document}